\documentclass{pas}
\usepackage{multirow}

\usepackage{float}
\setcitestyle{notesep={}}
\newcommand{\kms}{\,km\,s$^{-1}$}

\begin{document}


\lefttitle{Publications of the Astronomical Society of Australia}
\righttitle{S. Deeley et al.}

\jnlPage{1}{4}
\jnlDoiYr{2021}
\doival{10.1017/pasa.xxxx.xx}

\articletitt{Research Paper}
\title{The H$\alpha$ specific angular momentum of dwarf galaxies}

\author{
Simon Deeley$^{1}$\thanks{E-mail: s.deeley@uq.edu.au}, Sarah Sweet$^{1,2}$, Dilyar Barat$^{3}$, Matthew Colless$^{2,3}$, Luca Cortese$^{2,4}$ and ~Francesco D’Eugenio$^{5,6}$ 
}
\affil{
${}^1$School of Mathematics and Physics, University of Queensland, Brisbane, Queensland 4072, Australia\\
${}^2$ARC Centre of Excellence for All Sky Astrophysics in 3 Dimensions (ASTRO 3D)\\
${}^3$Research School of Astronomy and Astrophysics, Australian National University, Canberra, ACT 2611, Australia \\
${}^4$International Centre for Radio Astronomy Research (ICRAR), University of Western Australia, Crawley, WA 6009, Australia \\
${}^5$Kavli Institute for Cosmology, University of Cambridge, Madingley Road, Cambridge, CB3 0HA, United Kingdom\\
${}^6$Cavendish Laboratory - Astrophysics Group, University of Cambridge, 19 JJ Thomson Avenue, Cambridge, CB3 0HE, United Kingdom
}
\corresp{S. Deeley, Email: s.deeley@uq.edu.au}


\history{(Received xx xx xxxx; revised xx xx xxxx; accepted xx xx xxxx)}


\begin{abstract}
The relationship between a galaxy's specific angular momentum $j$ and its mass $M$, parameterised by $j \propto M^\alpha$ and known as the Fall relation, has emerged as a fundamental scaling relation reflecting key physical and morphological properties of galaxies. This relation has been well studied for galaxies with masses above $10^{9}$\,$M_{\odot}$. However, whether or not it holds for the low-mass dwarf galaxies, especially given their varied morphologies, remains uncertain. Here we use H$\alpha$ observations of 49 star-forming dwarf galaxies from the SH$\alpha$DE survey, as well as 20 high-mass `control' galaxies, to investigate the stellar $j_{*}$--$M_{*}$ relation down to masses below $10^{6}$\,M$_{\odot}$. We find that the star-forming dwarf galaxies follow the same $j_{*}$--$M_{*}$ relation as high-mass disk-like galaxies, with $\alpha = 0.53 \pm{0.4}$, demonstrating that the relation holds across 5 orders of magnitude in mass. We then select a matching sample from the IllustrisTNG cosmological simulation and create mock observations resembling the SH$\alpha$DE survey. We find that the simulated dwarf galaxy population follows the extrapolated $j_{*}$--$M_{*}$ relation with a flattening and significant scatter towards lower $j_{*}$ values. Following the evolution of these galaxies, we find that dwarf galaxies experience a gradual loss in $j_{*}$ with time, while high-mass galaxies experience a sudden jump in $j_{*}$ before settling into a stable state. This dynamical evolution leads to a redshift dependence in $\alpha$, with $\alpha = 0.45$ at $z = 2$ and 0.55 at $z = 0$. Despite the apparent simplicity in the present-day $j_{*}$--$M_{*}$ relation over a wide mass range, the evolution of $j$ leading to this relation is complex and dynamic. 
\end{abstract}



\maketitle

\section{Introduction}
\label{introduction}

One of the most important aspects in our understanding of galaxy evolution is how galaxies initially acquired their angular momentum and how this angular momentum evolved over time, leading to the diversity of galaxy structures we observe today. By our current understanding, in the early Universe gravitational forces from neighbouring over-densities created a tidal torque on each halo, acting to gradually increase their angular momentum as they continued to grow in mass through accretion and in size with the expanding Universe \citep{H..1949}. This continued until the inward gravitational force out-competed the rate of expansion and the halos began to collapse, at which point the strength of the tidal forces greatly reduced, effectively freezing in the halo's angular momentum ($J$) at that time. The time at which this `turnaround' occurred, and hence the angular momentum gained, was dependent on the halo's mass, leading to an expected relation between the halo's mass and its specific angular momentum $j \propto {\rm M}^{\alpha}$, where $j$ = $J/M$ and $M$ is the halo mass. \citet{1969ApJ...155..393P} analytically derived a value of 2/3 for the parameter $\alpha$, and additional analytical approaches were carried out by \citet{1970Afz.....6..581D} and \citet{1984ApJ...286...38W}. This was tested in numerical simulations by \citet{1979MNRAS.186..133E} and found to be in agreement with the expected relation. 

As the gas within the halo collapses further into the centre and forms the central galaxy, the forming galaxy and its stars should then (at least in an idealized closed-box scenario where the halo angular momentum is conserved) have the same angular momentum as the host halo. The mass-angular momentum of the dark matter halo would therefore be reflected in a mass-angular momentum relation of the central galaxies, which we are able to observe and measure. The first observational test of this relation was made by \citet{1983IAUS..100..391F}, who found that galaxies do indeed follow a $j_{*}$--$M_{*}$ relation with $\alpha = 0.55$, where $j_{*}$ and $M_{*}$ refer to the stellar component of $j$ and the stellar mass respectively. This relation is now commonly referred to as the Fall relation. As detailed below, a growing number of studies have since confirmed this relationship and explored how it varies with morphology and bulge/disk components.

Studies have revealed a fundamental connection between the $j$--$M$ relation and the morphology of galaxies. While the value of $\alpha$ remains largely independent of galaxy morphologies, the normalisation of this relation is found to be morphology-dependent. Spiral galaxies have the highest $j_{*}$ for a given mass, while elliptical galaxies have $j_{*}$ values a factor of 5 times lower \citep{2012ApJS..203...17R}. More generally, there is a clear trend of decreasing $j_{*}$ from late-type disk galaxies through lenticular galaxies to early-type elliptical galaxies \citep{2016MNRAS.463..170C}. There is also a strong related connection between $j_{*}$ and the bulge mass fraction; galaxies with a large bulge have lower $j_{*}$ values relative to those with smaller bulges \citep{2014ApJ...784...26O}. In fact, bulges and disks, when separated from each other, are each found to independently follow parallel $j_{*}$--$M_{*}$ relationships, with the bulges having lower $j_{*}$ relative to the disks (being closer in line with elliptical galaxies) \citep{2018ApJ...868..133F}.  The bulge type is also important, with galaxies that host rotating pseudobulges tending to have lower bulge fraction for a given $j_{*}/M_{*}$ than galaxies that host classical bulges \citep{2018ApJ...860...37S}. The fundamental relationship between a galaxy's angular momentum and its morphology has been proposed as an alternative, physically motivated way to classify galaxies in place of the visual component-based system widely used today \citep{2020MNRAS.494.5421S}. 

An additional physical property that has been found to be reflected in the $j$--$M$ relation is the galaxy's gas mass fraction. \citet{2021A&A...647A..76M} studied an amalgamated sample of 157 galaxies with neutral HI gas rotation curves and found $\alpha =$ 0.54, 1.02 and 0.6 for the stellar, gas and baryonic (star + gas) components respectively, finding that the stellar and baryonic $j$ is higher for galaxies with higher gas mass fractions. \citet{2021A&A...651L..15M} found that by incorporating the gas mass fraction, they were able to produce a tight 3D relationship between galaxy mass, gas mass fraction and specific angular momentum, with galaxies containing higher gas mass fractions featuring higher angular momenta. Using a sample of 559 galaxies observed through their HI emission, \citet{2022MNRAS.516.4043H} also found that the scatter in the $j_{*}$--$M_{*}$ relation is largely explained by the fraction of neutral gas in the galaxies, with galaxies containing a higher fraction of HI gas again having higher values of $j_{*}$. This observed gas-fraction relationship was found to be reproduced by both the EAGLE and IllustrisTNG simulations, with the two simulations being in close agreement with each other \citep{2023MNRAS.526..808H}. 

The $j_{*}$--$M_{*}$ relation obtained from observed galaxies has been successfully reproduced within various cosmological simulations. Using the Illustris TNG100 simulation, \citet{2022MNRAS.512.5978R} found a $j_{*}$--$M_{*}$ relation that agrees with observations down to masses of $10^{9}\,\rm M_{*} (M_{\odot})$, including the observed lower $j_{*}$ for ellipticals relative to spirals, and further concluded that the baryonic component retains 50--60 per cent of their host halo's angular momentum in spiral galaxies and 10--20 per cent in elliptical galaxies. \citet{2023NewA...9901964E}, using the SIMBA cosmological simulation, also found a $j_{*}$--$M_{*}$ relation consistent with a single power law down to galaxy masses of $10^{8}\,\rm M_{*} (M_{\odot})$, finding that galaxies with a higher mass of HI gas have higher angular momentum, in agreement with the observations detailed above. \citet{2017MNRAS.464.3850L} showed that the $j_{*}$--$M_{*}$ relation is reproduced within the EAGLE cosmological simulation. These simulations can therefore be used to investigate how the evolutionary history of galaxies impacts their final $j_{*}$ values.

\begin{figure*}
\begin{center}
\includegraphics[width=2\columnwidth]{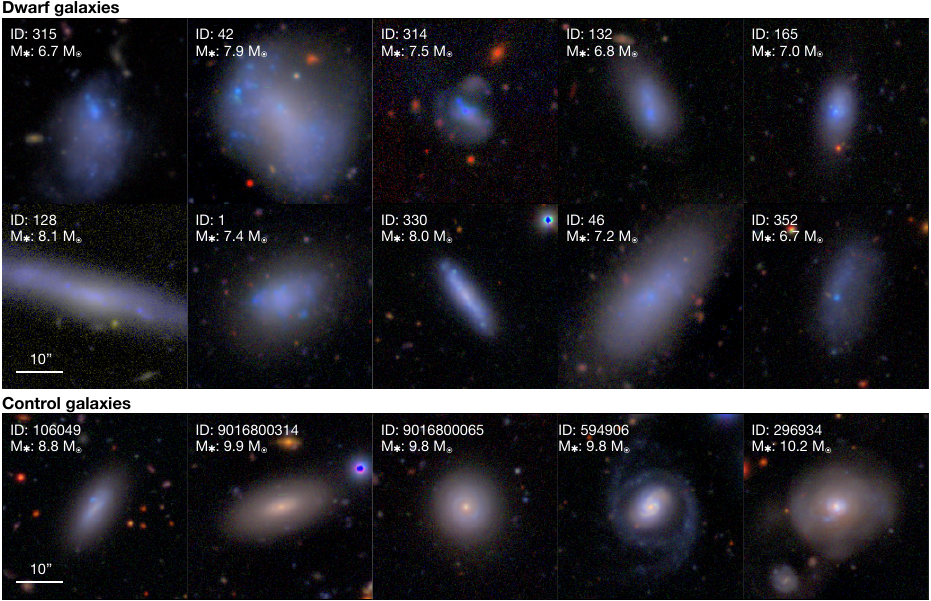}
   \caption{HSC imaging of dwarf galaxies (top two rows) and higher-mass control galaxies (bottom row). The dwarf galaxy sample includes a wide range of galaxy morphologies, from highly irregular and clumpy galaxies to thin, well-organised disks. The blue colours reflect the H$\alpha$ flux selection of these galaxies, and therefore their high star formation rates. The control sample, meanwhile, consists mainly of large disk galaxies. The scale bars in the lower left of each row indicates 10 arcseconds. Note that the control galaxies are located further away such that they are contained within the same instrumental field of view. }
 \label{HSC}
\end{center}
\end{figure*}

The final angular momentum of a galaxy may be significantly influenced by the timing of the initial build-up of its mass and stars. Using EAGLE, \cite{2016MNRAS.460.4466Z} showed that for galaxies that formed all their stars before the point at which the host halo begins to contract in size (the turnaround point), the stars subsequently lose their angular momentum due to merger activity during the continued build-up of the inner halo. Stars that form after this turnaround point do so from high-angular-momentum gas falling in from the halo's outskirts, resulting in a stellar disk with high angular momentum. 

It has been seen that subsequent galaxy evolution may further alter the amount of angular momentum measured in galaxies today. As already mentioned, elliptical galaxies have lower $j_{*}$ values and are believed to have formed through highly disruptive merger events. \cite{2018MNRAS.473.4956L} used the EAGLE simulation to investigate the impact of mergers on $j_{*}$, finding that both major and minor mergers can significantly alter a galaxy's angular momentum. Mergers with galaxies containing little gas (`dry mergers') caused a greater loss of $j_{*}$, while prograde gas-rich mergers were able to spin up the galaxy. In addition, they found that early quenching of star formation also results in a galaxy with lower $j_{*}$ and an elliptical morphology. Using the Illustris simulation, \cite{2015ApJ...804L..40G} also found major mergers to reduce a galaxy's angular momentum, and additionally found $j_{*}$ to be influenced by gas flows. Galactic outflows of gas, driven by outward pressure from star-forming regions, preferentially removes gas with lower angular momentum. This gas flows out into the outer halo where it can regain angular momentum from the halo's outskirts, before cooling and falling back in towards the galaxy. Meanwhile, an alternative pathway for galaxy quenching, namely the infall of galaxies into dense cluster environments, may leave the $j_{*}$--$M_{*}$ relation largely unaffected \citep{2021A&A...652A..10M}. Given the wide range of complex histories and environments experienced by each individual galaxy, it is still not well understood why they fall onto a tight relation between their mass and angular momentum, and why the slope of this relation is retained despite large changes in $j$ after dramatic morphological transformations. 

Whether or not the $j_{*}$--$M_{*}$ relation extends down to the dwarf galaxies has not yet been fully resolved. One of the most comprehensive observational studies into the Fall relation was carried out by \citet{2018A&A...612L...6P}, who looked at the stellar kinematics of 92 disk galaxies with masses between $10^{7}$ and $10^{11}\,\rm M_{*} (M_{\odot})$ and found a tight $j_{*}$--$M_{*}$ relation with $\alpha$ = 0.66. However, their sample included only six galaxies below $10^{8}\,\rm M_{*} (M_{\odot})$. Most studies of dwarf galaxies to date have focused on their HI components, due to the gas-rich nature of these systems and the extension of HI observations to larger radii. \citet{2017ApJ...834L...4B} measured the angular momentum of dwarf galaxies between $10^{6}$ and $10^{9}\,\rm M_{*} (M_{\odot})$, finding that they had higher angular momentum than expected from an extrapolation of the relation found for larger spirals. They argued that this may be due to the higher mass-to-light ratio of dwarf galaxies. \citet{2018MNRAS.479..228K} extended this to isolated dwarf galaxies located in voids and again found a higher degree of angular momentum for these galaxies. When combining with high-mass galaxies in previous work, they saw evidence for a transition in momentum at a mass of $10^{9.1}\,\rm M_{*} (M_{\odot})$, with galaxies below this threshold mass exhibiting higher angular momentum. \citet{2017MNRAS.467.3856C} measured $j$ for 5 dwarf galaxies between $10^{7.5}$ and $10^{9}\,\rm M_{*} (M_{\odot})$ and again found a higher $j$ then predicted from the extrapolation. Both \citet{2017MNRAS.472.4551E} and \citet{2021A&A...647A..76M} meanwhile found no divergence from the relation for galaxies between $10^{7}$ and $10^{11}\,\rm M_{*} (M_{\odot})$. Specifically, \citet{2021A&A...647A..76M} tested for a broken power law in the relation and found that the sample was better explained using a single power law. Alternatively, any change in the $j_{*}$--$M_{*}$ relation among dwarf galaxies may arise from an offset in the absolute values rather than a change in slope, as is the case for the spiral and elliptical galaxies. A larger sample of dwarfs over a wide mass range would be needed to reveal such an offset. 

Most of these studies assume that dwarf galaxies are pure rotating disks, and while hydrogen traces well the smooth disk out to large radii, it is not sensitive to small-scale kinematic features within the star-forming optical extent. Here we use an alternative tracer of specific angular momentum in dwarf galaxies. With many dwarf galaxies being gas-rich star-forming systems, the bright H$\alpha$ emission line offers an ideal alternative to accurately trace their specific angular momenta in these otherwise faint galaxies. H$\alpha$ emission has been used previously to study the $j_{*}$--$M_{*}$ relation of higher-mass galaxies; for example, \citet{2016MNRAS.463..170C} used the SAMI Galaxy Survey to derive the $j$--$M$ relation for both the stellar and H$\alpha$ component. However, even the relatively high spectral resolution of the SAMI spectroscopy ($R = 4263$ at $H\alpha$, corresponding to a velocity resolution of 30\kms) is insufficient to accurately measure the velocity dispersion in low-mass dwarf galaxies \citep{2020MNRAS.498.5885B}, except for the few, highest-quality observations \citep{2017MNRAS.470.4573Z}. To address this limitation, here we use spatially-resolved spectroscopy targeting warm ionised gas emission with high spectral resolution to measure $j_{*}$ in a sample of low-mass galaxies, testing the extension of the $j_{*}$--$M_{*}$ relation into the dwarf galaxy regime. Throughout this paper we assume a cosmological constant plus cold dark matter cosmology ($\Lambda$CDM) with $\Omega _{M}$ = 0.3, $\Omega _{\Lambda}$ = 0.7 and H$_{0}$ = 70\kms, and a Chabrier initial mass function \citep{2003PASP..115..763C}.

\section{Data}

In this work we combine an observational survey of star-forming dwarf galaxies with a sample taken from the IllustrisTNG-50 cosmological simulation in order to further explore the extension of the Fall relation to low-mass galaxies. Below we outline the galaxy samples used for each of these two components.

\subsection{Observational sample}

To investigate the angular momentum of dwarf galaxies in the local Universe we use data from the Study of H$\alpha$ from Dwarf Emissions (SH$\alpha$DE) survey \citep{2020MNRAS.498.5885B}. The SH$\alpha$DE survey was designed to extend existing large integral field unit (IFU) surveys such as SAMI, which target bright high-mass galaxies, into the low-mass dwarf galaxy regime. The aims of the survey were to investigate the extension of scaling relations into the low-mass regime, the accretion of angular momentum, the asymmetry of dwarf galaxy kinematics and the dynamical impact of star formation.

\begin{figure}
\includegraphics[width=1\columnwidth]{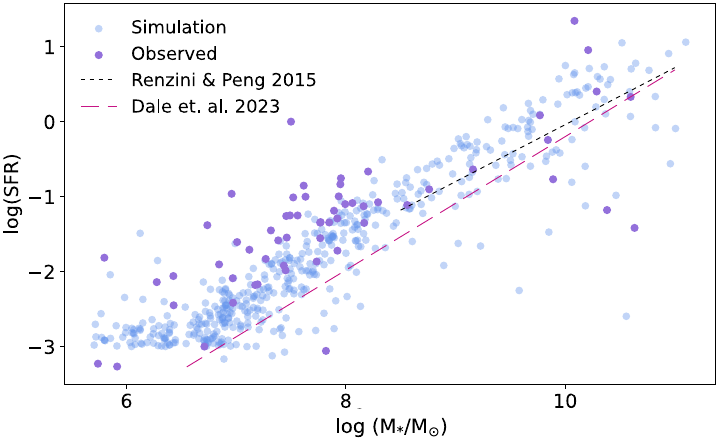}
   \caption{The star-forming sequence of our simulation sample (blue) compared to our observational sample (purple). For comparision, the star-forming main sequence found by \citet{2015ApJ...801L..29R} for galaxies above 10$^{8.5}$ M$_{\odot}$ and for local-volume galaxies down to 10$^{6}$ M$_{\odot}$ \citep{2023AJ....165..260D} is shown. Below 10$^{6.5}$ M$_{\odot}$, the sample becomes biased towards higher star formation rates due to our selection limits.}
 \label{sfs}
\end{figure}

The complete details of the SH$\alpha$DE survey are presented in \citet{2020MNRAS.498.5885B}. In summary, the SH$\alpha$DE survey targeted 49 dwarf galaxies within a stellar mass range of $10^{5} \leq M_{*} ({\rm M}_{\odot}) \leq 10^{8.5}$ and with apparent angular sizes in the range $1.2 \leq R_d ({\rm arcsec}) \leq 11$. To ensure their kinematics could be accurately measured through their H$\alpha$ emission, their total H$\alpha$ fluxes were required to be above $5\times 10^{-16}$\,ergs\,s$^{-1}$\,cm$^{-2}$\,\AA\,arcsec$^{-2}$. This requirement resulted in the sample consisting of star-forming dwarf galaxies; Figure~\ref{sfs} shows the location of this sample (purple points) compared to established relationships between mass and star formation rate. 

As part of the SH$\alpha$DE survey, a sample of 20 'control' galaxies within the stellar mass range of $10^{8.5} \leq M_{*} ({\rm M}_{\odot}) \leq 10^{10.5}$ (selected from the SAMI survey) were also included, allowing us to verify the consistency of our methods with previous work and to test with a consistent sample whether scaling relations found amongst the high-mass galaxies can be continuously extended down to the dwarf galaxy population. 

\begin{figure*}
\begin{center}
\includegraphics[width=2\columnwidth]{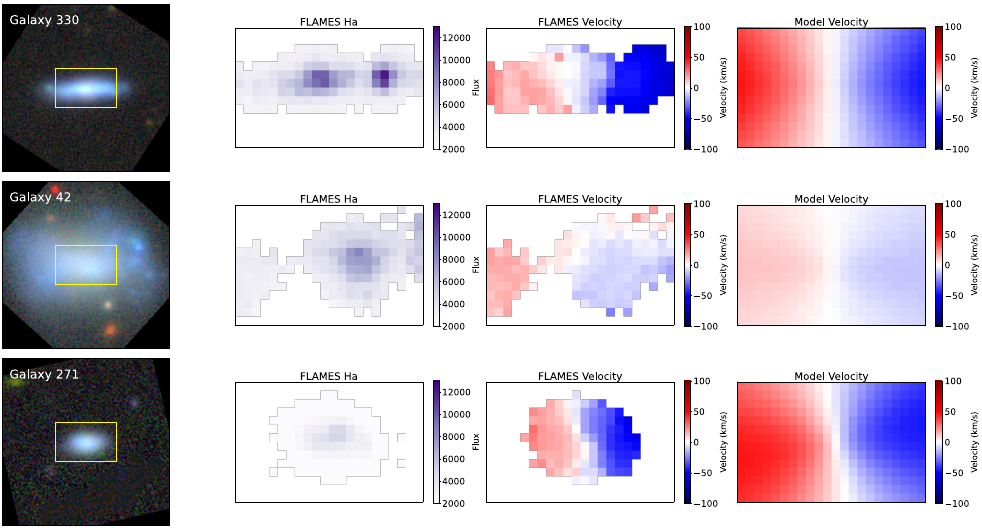}
   \caption{Examples of the SH$\alpha$DE kinematic maps and the corresponding velocity field models. The left column shows optical images taken by the Dark Energy Legacy Survey, with the FLAMES footprint shown by the yellow box. The second and third columns display the H$\alpha$ emission and the H$\alpha$-derived velocity map respectively, and the final column shows the velocity field produced by the best-fitting rotating-disk model of the galaxy.}
 \label{kinematics}
\end{center}
\end{figure*}

All 69 galaxies were observed with the FLAMES/GIRAFFE IFU instrument on the Very Large Telescope using the GIRAFFE spectrograph in `Argus' mode, a single IFU consisting of 22 x 14 microlenses, with each lens covering 0.52 arcsec on the sky. This results in a $11.4\times7.3\, {\rm arcsec}^{2}$ array of 308 spectra. From the measured spectrum within each spaxel, pPXF \citep{2023MNRAS.526.3273C} was used to measure the velocity and velocity dispersion from the H$\alpha$ emission line, resulting in 2D maps of the velocity and velocity dispersion for each galaxy. 

14 galaxies have been imaged by the Hyper-Suprime Cam Subaru Strategic Program survey \citep{2018PASJ...70S...4A}, allowing a detailed view of their morphology. We used the $g$, $i$ and $r$-band images to construct red-green-blue colour images for these galaxies, eight of which are shown in Figure~\ref{HSC}. These examples highlight the range of morphologies in our sample, which includes very disky galaxies as well as highly irregular and clumpy galaxies. Owing to the selection of galaxies with high H$\alpha$ flux, all these galaxies are blue star-forming galaxies, with many featuring prominent clumps and regions of ongoing star formation. Four of the higher-mass control galaxies are also shown in Figure~\ref{HSC}. The full sample of galaxies has been imaged by the DESI Legacy Imaging Surveys \citep{2019AJ....157..168D}, which we have used to visually classify the morphology of each galaxy as either disk-like, irregular, or compact (i.e.\ having either a very compact or unresolvable structure). 

\begin{figure}
\includegraphics[width=1\columnwidth]{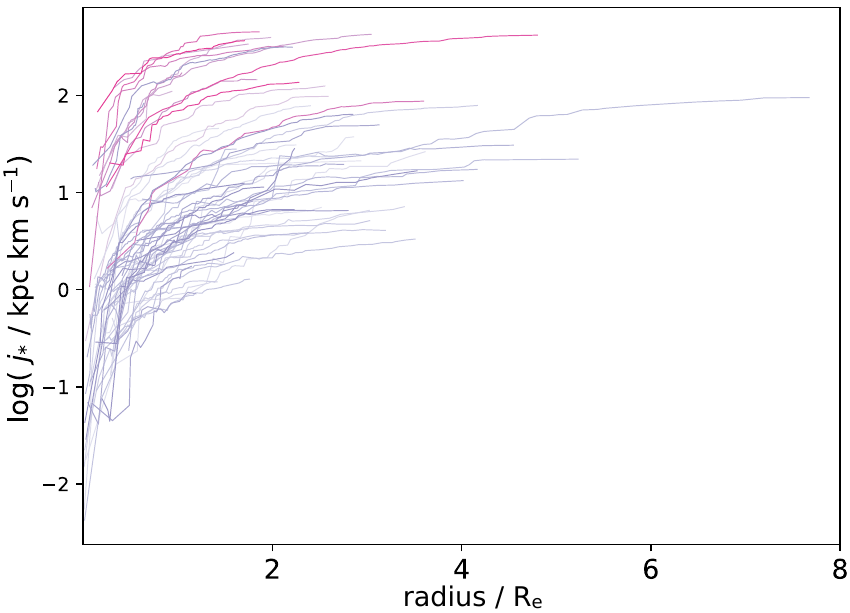}
   \caption{$j_{*}(<r/R_e)$ as a function of radius for the dwarf (blue) and control (red) galaxies. The radial profiles approach a constant value towards the edge of our observed limits in most cases, showing that our observations extend far enough over the galaxies to capture the bulk of their total angular momentum.}
 \label{cumulative}
\end{figure}

\begin{figure*}
\begin{center}
\includegraphics[width=1.8\columnwidth]{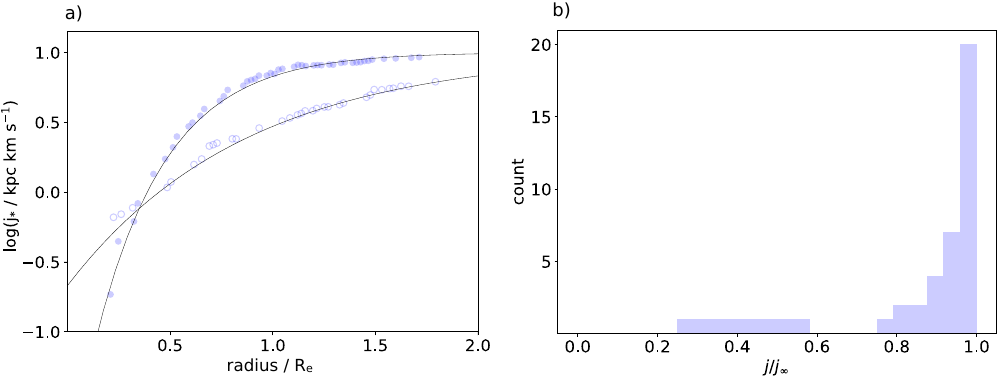}
   \caption{Convergence of the cumulative $j_{*}(<r/R_e)$--$R$ profiles for our observational sample. a) displays the cumulative $j_{*}$ profile for a converging (ID: 1, filled points) and non-converging (ID: 322, open points) galaxy, along with their respective best-fit exponential curve. The exponential function provides a good description of the profile and demonstrates how well each profile has converged on the total $j_{*}$ value. b) shows the distribution of $j_{*,max}/j_{*,\infty}$, illustrating that the majority of our galaxies are approaching convergence.}
 \label{convergence_eg}
\end{center}
\end{figure*}

\subsection{Simulation sample}

For the simulation comparison, we used the IllustrisTNG cosmological simulation \citep{2019ComAC...6....2N,2018MNRAS.480.5113M,2018MNRAS.477.1206N,2018MNRAS.475..624N,2018MNRAS.475..648P,2018MNRAS.475..676S}. We specifically use the smallest-volume, highest-resolution TNG50 simulation \citep{2019MNRAS.490.3234N,2019MNRAS.490.3196P,2019ComAC...6....2N} in order to allow us to more finely resolve low-mass dwarf galaxies. The simulation's volume is 51.7\,Mpc$^{3}$, allowing for the study of galaxies in a wide range of environments with the largest structures being of similar size to the Virgo galaxy cluster. The simulation features particle resolutions for baryonic and dark matter particles of $8.5\times 10^{4}$\,M$_{\odot}$ and $4.5\times 10^{5}$\,M$_{\odot}$ respectively.

The crucial benefit in using TNG50 for this work is the ability to study dwarf galaxies and their acquisition of angular momentum within a broader cosmological context, and to consistently compare these galaxies to high-mass disk galaxies in the same context. In particular, TNG50 offers the highest resolution currently available while maintaining such a cosmologically-relevant context. We note that we are approaching the limits of what the simulation can reliably resolve, meaning that resolution effects may have an influence on our findings. For example, potential influences on galaxy structure have been identified by \citet{2025A&A...699A..12C}, namely a flattening of the mass-radius relation below $10^{8}\,\rm M_{*} (M_{\odot})$. However, as we will further detail below, we also note that the dynamical transitions we see within our simulation results are consistent with higher resolution zoom-in simulations \citep{2025arXiv250800991B}, and we find consistent results within the lower-resolution TNG100 simulation. While these comparisons suggest TNG50 can still produce dynamical processes for such low-mass galaxies, we stress that potential resolution effects should be kept in mind through the rest of this paper. 

From this simulation we selected a sample of 512 galaxies in the stellar mass range $10^{6} \leq M_{*}({\rm M}_{\odot}) \leq 10^{10.5}$. Due to the increased computational time for deriving observational maps for high-mass galaxies and the focus of this work on testing the Fall relation within the dwarf galaxy regime, we have over-sampled below $10^{8}$M$_{\odot}$ relative to the higher-mass galaxies. We then applied a further selection cut based on star formation rate (SFR), reflecting the minimum H$\alpha$ emission requirement of our observational sample. This minimum effectively corresponds to a lower limit in star formation \citep{1998ARA&A..36..189K}, with all except four of our observed galaxies having $\log_{10}({\rm SFR}) > -3$; we therefore use this SFR as a lower limit for our simulation sample. This fixed SFR limit results in the preferential removal of galaxies with low star formation at lower masses relative to higher masses. Investigating the mass-SFR distribution of our sample shows that this begins to bias the simulation sample below $10^{7} \leq \,\rm M_{*} (M_{\odot})$. The mass-SFR distribution of the  simulation sample is shown in Figure~\ref{sfs} as the blue points. 

\section{Methods}

\subsection{\texorpdfstring{SH$\alpha$DE}{SHADE}}

To calculate $j_{*}$ we follow the methods of \citet{2020MNRAS.494.5421S}, combining robust observational kinematic and surface density maps with a model extrapolation to 3 effective radii. 

\subsubsection{Kinematics}

Firstly, in order to correctly model the velocity field, we removed spaxels with high uncertainties or unphysical values while avoiding the removal of spaxels with valid and important information. We began by removing all spaxels with a signal-to-noise ratio $\leq$5. To remove remaining spaxels with large uncertainties without significantly reducing the number of available pixels for analysis, we also removed those with a line-of-sight velocity uncertainty greater than 30\kms. After these initial cuts, isolated high-velocity spaxels remained in a number of kinematic maps, many of which had velocity values that deviated significantly from the rest of the map. In addition, there were a number of spaxels with velocity signs opposite to all neighboring spaxels. We tested two methods to remove these outliers. The first method involved sigma clipping, where we removed all spaxels with velocities more than 2.5 standard deviations from the mean. While this successfully removed all isolated and contrary spaxels, it had the undesirable side-effect of removing high-velocity clumps that were clearly real features of the galaxy's velocity field. The second method involved removing all spaxels with no neighbours or with all surrounding spaxels having the opposite sign. While this left some spaxels with highly divergent velocities, it retained the important features of the velocity fields. We carried out the analysis using both methods and found that, aside from some shifts in $j$ for individual galaxies, the overall $j$--$M_{*}$ relations were virtually unchanged between the two approaches; for the remainder of this paper we present results using only the second method. 

We then deprojected the velocity fields and created a kinematic model of each galaxy assuming a thin, axisymmetric exponential disk. We use the continuum flux map to determine the inclination and the line-of-sight velocity map to find the kinematic centre and position angle, from which we determined the deprojected radius and velocity of each spaxel. We then fitted the radial velocity profile with an exponential profile and from this created a model velocity field. We test the assumption of a thin kinematic disk and our method of deprojection in Section~\ref{thin-disk test} by (alternatively) deprojecting the line-of-sight velocities using optically-derived ellipticities and varying the assumed disk thickness (parametrised by $q_{0}$), finding that our results remain consistent when using this alternative approach.

\subsubsection{Calculation of \texorpdfstring{$j_{*}$}{j}}

To arrive at a measure of the stellar angular momentum, which requires both a measure of the stars' motions and mass distribution, we use the kinematics derived from the H$\alpha$ emission-line maps (see \citet{2020MNRAS.498.5885B} for how these maps were created) and assume that the stellar mass distribution follows the galaxy's continuum optical flux. While reflecting the dynamics of younger stellar populations, we note that, in using the kinematics of star-forming regions, this approach may result in $j$ values that vary from the broader stellar population, particularly an older population that may have formed under different dynamical circumstances and subsequently been influenced by evolutionary processes such as mergers. This should be kept in mind when interpreting the results of this paper and in making comparisons to other work.

In modelling the continuum-derived stellar mass distribution, we assume the distribution to take the form of an exponential disk and normalise the total continuum flux by the galaxy's total mass. Using the deprojected velocities derived from the kinematic modelling above, we then go through all $N$ spaxels and calculate the observational $j_{*}$ by
\begin{equation}
j_{*} = \frac{\sum_{i=0}^{N} M_i v_i r_i}{\sum_{i=0}^{N}M_{i}},
\label{equation}	
\end{equation}

where $M_i$ is the mass, $r_i$ is the radius, and $v_i$ is the deprojected line-of-sight velocity of the $i^{\rm th}$ spaxel. In addition, we also calculate $j_{*}$ for all galaxies using the velocity field of the best fitting disk model, restricted to within 3 effective radii ($R_e$); we call this $j_{*\rm model}$. Finally, we use the model velocity field to fill in missing spaxels in the observational map to create a combined model + observational velocity map, and calculate a combined $j$ value ($j_{*\rm total}$) (again out to a limit of 3$R_e$ for all galaxies). While including the model mass and velocity field for galaxies that are approaching convergence systematically increases the absolute value of $j_{*}$ due to the filling in of spaxels which otherwise have no observational data, we find that our main conclusions are not significantly affected by the choice of $j_{*\rm obs}$, $j_{*\rm model}$ or $j_{*\rm tot}$. To prioritise observational data while compensating for the varying extent to which each galaxy has reliable measurements, we therefore focus on $j_{*\rm tot}$ for the remainder of this paper and refer to this simply as $j_{*}$. 

To quantify the $j_{*}$--$M_{*}$ relation, following previous studies we fit the expected relationship $j_{*} \propto M^{\alpha}$ using the form
\begin{equation}
\log j_{*} = \alpha \left( \log(M_{*}/M_{\odot}) - 11 \right) +\beta
\label{logj-logM}	
\end{equation}
where $\alpha$ and $\beta$ are the free parameters to be fitted. 

\subsubsection{Convergence of \texorpdfstring{$j_{*}$}{j}}

The cumulative $j_{*}$ typically increases with increasing radius. If the specific angular momentum is still increasing beyond the region of the galaxy covered by the observations, the galaxy's total $j_{*}$ may be higher than the value we can determine from the observed velocity map. In addition to the limitations imposed by the instrumental field of view, the distribution of star formation can be patchy and may differ from the distribution of stars, so H$\alpha$ velocity measurements may not fully trace a galaxy's velocity field. To check for these biases and determine if our calculated values of $j_{*}$ correspond to the total $j_{*}$ of the galaxy, we calculated the cumulative value of $j_{*}$ as the radial limit is increased from zero out to the edge of the field of view, and looked to see if $j_{*}$ converges to a constant value within the radius of our measurements. 

The cumulative specific angular momentum as a function of radius for all galaxies in our sample is shown in Figure~\ref{cumulative}. In the majority of cases, $j_{*}$ appears to be converging to a constant value within the limits of our observations. To check this quantitatively, we followed a similar procedure to \citet{2021A&A...647A..76M}, though in our case we found the cumulative $j_{*}$ profiles to be well represented by an exponential function; see for example Figure~\ref{convergence_eg}. Taking the value of the fitted function at infinity as the $j_{*}$ value each profile is converging towards ($j_{*,\infty}$), we used the ratio of the total $j_{*}$ value attained within our measurements with ($j_{*,\infty}$) as a measure of how much each profile has converged. We found that 59 (85 percent) of our galaxies attain at least 80 percent of their extrapolated ($j_{*,\infty}$) value within our observational footprint. 

\begin{figure*}
\begin{center}
\includegraphics[width=2\columnwidth]{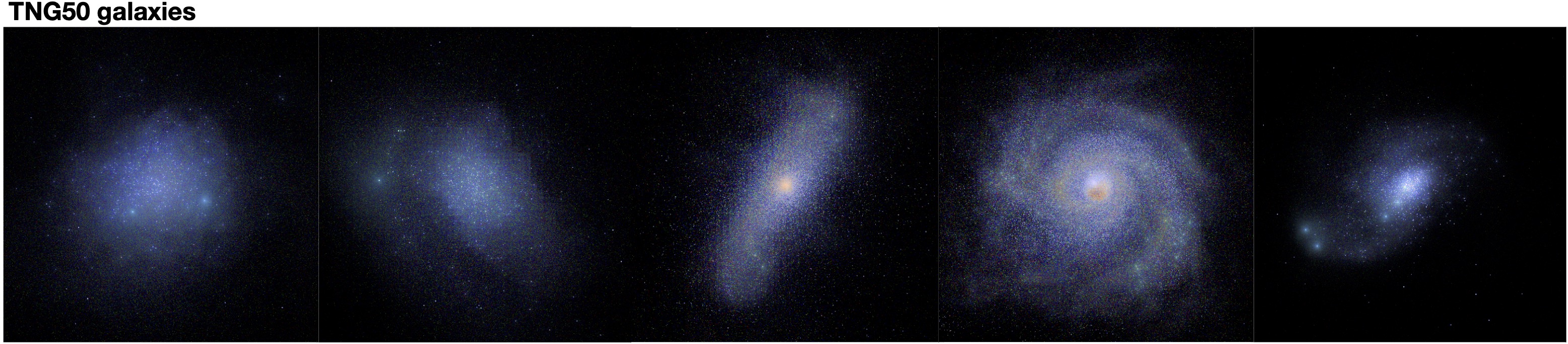}
   \caption{Optical images of example simulated dwarf and low-mass spiral galaxies created using the radiative transfer code SKIRT . The images demonstrate the range of morphologies in the simulated galaxies, which resemble the variety of morphologies in our observational sample (see Figure~\ref{HSC}).}
 \label{sim_images}
\end{center}
\end{figure*}

\begin{figure*}
\begin{center}
\includegraphics[width=2.0\columnwidth]{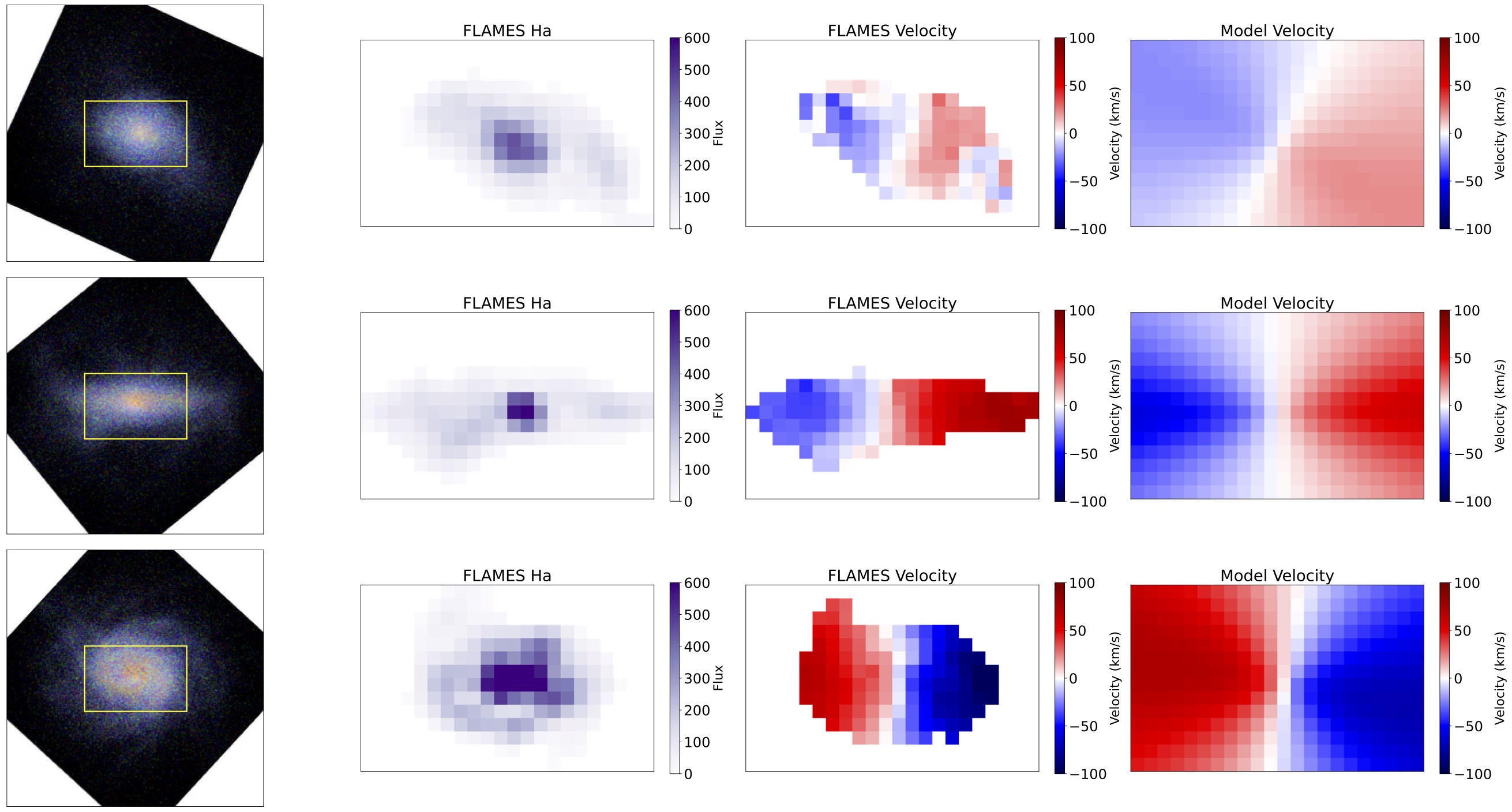}
   \caption{Examples of observationally-equivalent flux and kinematic maps for the simulated galaxies. The maps were made to mimic the SH$\alpha$DE survey, showing the H$\alpha$ flux and line-of-sight velocity over a 14$\times$22 spaxel window resembling the FLAMES footprint. These maps were then analysed using the same methods as for the observational sample, resulting in the modelled kinematic maps shown in the right-hand column. The optical images on the left-hand-side were created using the radiative transfer code SKIRT.}
   \label{sim_kin}
   \end{center}
  \end{figure*}

\subsection{Simulation}

\subsubsection{Optical images}

We first check that the morphology of our simulated galaxies reproduces the observational sample they are compared to. We use the radiative transfer code SKIRT \citep{2015A&C.....9...20C} to simulate light being emitted from the stars and travelling through the galaxy interstellar medium before reaching a hypothetical VLT observatory. The galaxy is placed at a distance such that the FLAMES footprint covers 3 effective radii. We largely follow \citet{2019MNRAS.483.4140R}, who created SKIRT images for higher-mass IllustrisTNG galaxies. We employ the \citet{2003MNRAS.344.1000B} Spectral Energy Distributions (SEDs) to model the emission spectrum of the stellar particles, selecting the Chabrier Initial Mass Function \citep{2003PASP..115..763C}. Star particles which are younger than 10 million years are considered to be star-forming regions, with a star formation rate equal to the mass of the stellar particle divided by 10 million years. The emission from these particles is modelled separately using the MAPPINGS-III SED \citep{2008ApJS..176..438G}, which specifically models star forming regions. Examples from the simulation sample are shown in Figure~\ref{sim_images}.  

\subsubsection{Mock observational maps}

For the simulation sample, we first calculated the true $j$ value separately for the star, gas and dark matter particles within 1, 2, 3 and 5\,$R_e$, and finally for all particles in the galaxy's halo. This was done by calculating the angular momentum of each individual particle around the centre of the galaxy and summing them to find the total angular momentum, and dividing this by the total mass of all particles used in the calculation. We also determined $j$ for the gas that was actively undergoing star formation to see how it is related to the $j_{*}$ of the resulting stars. This was carried out for all 100 snapshots of the simulation (from the time when the galaxy's progenitor halo first formed stars), resulting in the full time evolution of $j$ for each component of the galaxy. 

While the simulation has the great advantage of allowing us to calculate the galaxy's true angular momentum directly, this needs to be tied back to our observations. To this end, following the methods of \citet{2021MNRAS.508..895D}, we recreated our SH$\alpha$DE observations for the simulated galaxies so that we could measure $j_{*}$ in the same way as we did for the observed galaxies. This serves both as a test that our observational methods are able to recover the intrinsic $j_{*}$ of the galaxy and as a test that the simulation reproduces the distribution of angular momentum that we observe. For this we use the projection of the galaxy's particles onto the $x$--$y$ spatial plane (giving effectively a random projected orientation for each galaxy) and designate the $z$-axis as the line-of-sight direction to the observer. Firstly we fit concentric ellipses to the g-band flux distribution of the stellar particles in order to find the projected orientation of the galaxy, and hence the angle of its semi-major axis. We then overlay the FLAMES instrument footprint (consisting of 14 by 22 spaxels) onto the galaxy, angled so that its lengthwise direction is aligned with the galaxy's longest axis, and re-scale the galaxy's projected size so that the FLAMES window covers 3 times the galaxy's effective radius. 

Since our observed velocities are measured using the H$\alpha$ emission from star-forming gas, we use the star-forming gas particles in the simulated galaxies for the velocity calculations. IllustrisTNG does not model H$\alpha$ emission directly, however H$\alpha$ emission is closely related to the star formation rate, since it is the bright young stars ionizing their surrounding gas that leads to this emission. We therefore use the star formation rate of each gas particle to derive its H$\alpha$ emission, using the relation in \citet{1998ApJ...498..541K}. Even though this calibration assumes an IMF and solar metallicity, for our purposes here the absolute value of the H$\alpha$ emission is not important. The velocities of the gas particles are then weighted by their H$\alpha$ emission, and gas particles with no H$\alpha$ emission (which would be invisible to our observations) are left out of the calculation. 

\begin{table}
\renewcommand{\arraystretch}{1.01}
\setlength{\tabcolsep}{4pt}
\begin{center}
\vspace*{14pt}
\caption{Observational sample, with (a)~galaxy SH$\alpha$DE ID number; (b)~stellar mass; (c)~effective radius; (d)~$j_{*}$ calculated from the deprojected observational data; (e)~$j_{*}$ calculated from the model; and (f)~$j_{*}$ calculated from the combination of the observational data and model, where the model is used to fill in missing spaxels in the observational data. Columns (d), (e) and (f) are given in units of log($j$\,/\,kpc\,km\,s$^{-1}$).} 
\begin{tabular}{lccccc}
\hline
(a) & (b) & (c) & (d) & (e) & (f) \\
ID  &  log\,M$_{*}$/M$_{\odot}$  &  R$_e$\,(kpc)  &  log\,$j_{*,\rm obs}$  &  log\,$j_{*,\rm mod}$  &  log\,$j_{*,\rm tot}$  \\
\hline
1  &  7.4  &  0.9  &  0.87  &  1.01  &  1.01  \\
2  &  7.6  &  1.6  &  1.12  &  1.73  &  1.73  \\
9  &  7.9  &  2.5  &  1.89  &  2.07  &  2.07  \\
42  &  7.9  &  1.2  &  0.69  &  0.73  &  0.74  \\
46  &  7.2  &  0.7  &  0.58  &  0.94  &  0.94  \\
55  &  7.3  &  0.6  &  0.09  &  0.56  &  0.56  \\
56  &  7.8  &  1.2  &  0.97  &  1.02  &  1.03  \\
59  &  7.3  &  0.7  &  1.36  &  1.67  &  1.68  \\
64  &  6.3  &  0.9  &  0.59  &  0.76  &  0.77  \\
128  &  8.1  &  2.4  &  1.66  &  1.87  &  1.88  \\
132  &  6.8  &  0.8  &  0.62  &  0.82  &  0.84  \\
136  &  8.0  &  1.1  &  0.8  &  1.02  &  1.02  \\
137  &  7.9  &  1.0  &  1.99  &  2.2  &  2.21  \\
148  &  7.8  &  1.0  &  1.28  &  1.88  &  1.88  \\
151  &  8.2  &  1.2  &  1.49  &  1.72  &  1.72  \\
152  &  7.5  &  1.9  &  1.55  &  2.13  &  2.14  \\
165  &  7.0  &  0.7  &  0.72  &  1.09  &  1.08  \\
171  &  7.5  &  0.9  &  0.83  &  1.09  &  1.1  \\
194  &  7.0  &  1.2  &  0.7  &  0.86  &  0.87  \\
200  &  6.4  &  0.4  &  0.59  &  0.64  &  0.63  \\
218  &  7.4  &  1.0  &  1.07  &  1.3  &  1.3  \\
231  &  7.8  &  0.6  &  0.87  &  1.37  &  1.37  \\
260  &  5.9  &  0.4  &  0.53  &  0.61  &  0.61  \\
271  &  8.3  &  1.4  &  1.81  &  2.07  &  2.08  \\
283  &  8.2  &  1.7  &  0.64  &  0.74  &  0.73  \\
284  &  5.8  &  0.6  &  0.26  &  0.65  &  0.64  \\
286  &  7.2  &  0.7  &  -0.06  &  0.38  &  0.38  \\
288  &  7.7  &  0.8  &  0.56  &  1.29  &  1.29  \\
311  &  8.2  &  1.7  &  1.62  &  1.95  &  1.97  \\
314  &  7.5  &  2.0  &  1.38  &  1.67  &  1.67  \\
315  &  6.7  &  0.7  &  1.15  &  1.26  &  1.29  \\
319  &  7.5  &  1.5  &  1.3  &  1.67  &  1.68  \\
320  &  7.5  &  0.9  &  1.24  &  1.54  &  1.54  \\
322  &  7.0  &  1.0  &  0.92  &  1.3  &  1.31  \\
323  &  7.6  &  2.8  &  1.34  &  1.72  &  1.72  \\
327  &  7.9  &  1.8  &  1.5  &  1.77  &  1.78  \\
330  &  8.0  &  1.3  &  1.9  &  2.21  &  2.22  \\
331  &  7.9  &  1.1  &  1.21  &  1.62  &  1.62  \\
343  &  6.4  &  0.6  &  0.83  &  0.83  &  0.89  \\
344  &  7.1  &  1.8  &  1.49  &  1.74  &  1.76  \\
352  &  6.7  &  0.6  &  0.24  &  0.98  &  0.99  \\
424  &  10.1  &  7.2  &  2.51  &  2.72  &  2.72  \\
431  &  7.0  &  0.7  &  0.4  &  0.74  &  0.74  \\
433  &  7.6  &  1.1  &  1.43  &  1.41  &  1.46  \\
435  &  7.8  &  0.9  &  0.61  &  0.98  &  0.98  \\
469  &  7.5  &  0.7  &  0.78  &  0.84  &  0.91  \\
496  &  5.7  &  0.2  &  0.12  &  0.44  &  0.45  \\
520  &  7.8  &  0.8  &  1.7  &  2.08  &  2.08  \\
296934  &  10.2  &  5.0  &  2.64  &  2.98  &  2.96  \\
106049  &  8.8  &  2.3  &  2.0  &  2.24  &  2.26  \\
319150  &  8.6  &  1.5  &  0.88  &  1.25  &  1.25  \\
511921  &  9.2  &  1.6  &  1.94  &  2.25  &  2.25  \\
594906  &  9.8  &  2.4  &  2.62  &  2.7  &  2.73  \\
9008500356  &  10.4  &  5.4  &  2.64  &  3.2  &  3.2  \\
9011900128  &  9.5  &  3.6  &  2.15  &  2.47  &  2.47  \\
9016800065  &  9.8  &  4.8  &  2.16  &  2.53  &  2.52  \\
9016800314  &  9.9  &  4.5  &  1.95  &  2.72  &  2.72  \\
9091700123  &  9.7  &  3.5  &  2.11  &  2.31  &  2.32  \\
9091700137  &  10.7  &  6.1  &  2.66  &  2.89  &  2.89  \\
9091700444  &  10.4  &  4.6  &  2.6  &  2.92  &  2.92  \\
9239900178  &  10.6  &  4.6  &  2.08  &  2.7  &  2.7  \\
9239900182  &  10.6  &  4.8  &  2.57  &  2.9  &  2.9  \\
9239900237  &  10.3  &  5.3  &  2.65  &  2.9  &  2.91  \\
9239900370  &  10.0  &  4.4  &  2.24  &  3.15  &  3.15  \\
9388000269  &  10.5  &  4.2  &  2.52  &  2.92  &  2.91  \\
\hline
\end{tabular}
\end{center}
\end{table}

To then calculate a line-of-sight velocity and velocity dispersion for each spaxel of the FLAMES footprint, we firstly account for the point-spread function (PSF) of the observations by creating a Gaussian function centred on that spaxel, with a full-width-half-maximum corresponding to the SH$\alpha$DE average PSF of 0.88\,arcsec. We then create a weighted histogram of velocities (in the $z$ direction) using the value of that Gaussian at the location of each gas particle (in addition to the particle's H$\alpha$ emission) as the weight. The central velocity of the resulting histogram then corresponds to the measured line-of-sight velocity, and the spread (standard deviation, $\sigma$) corresponds to the velocity dispersion. This is equivalent to measuring the central velocity and width of the H$\alpha$ emission line. The same Gaussian is also used to find the weighted sum of the H$\alpha$ emission in order to find the H$\alpha$ flux of that spaxel, and it is used to weight the luminosities of stellar particles to determine the stellar flux. Finally, random noise is added to the velocity map in relation to the strength of the H$\alpha$ emission for each spaxel. 
 
Once we have the stellar flux, H$\alpha$ emission, and line-of-sight velocity maps for each galaxy, we can then follow the same process used to find $j_{*}$ for the galaxies observed by SH$\alpha$DE, allowing us to make a direct comparison to the observed values and link these to the intrinsic angular momentum and evolutionary history of these galaxies. Figure~\ref{sim_kin} shows examples of the resulting H$\alpha$ flux, velocity field, and modelled velocity field for three galaxies; these can be compared to the observed example shown in Figure~\ref{kinematics}. 

\subsubsection{Galaxy histories}

We then followed the evolution of each galaxy's angular momentum, to gain insight into how this relation arises and evolves. At each snapshot in the galaxy's history, we calculate the intrinsic $j_{*}$ of the star and gas particles in the same way as above (we do not derive observational-equivalent $j_{*}$ values at every snapshot due to the time-intensive nature of these calculations, instead using the intrinsic $j_{*}$ calculated directly from the particles). Following the same methods as \citet{2021MNRAS.508..895D}, we use the sublink halo trees to track and record merger events, including minor mergers (with a mass ratio between the progenitor galaxies of between 1:10 and 1:3) and major mergers (with a mass ratio greater than 1:3). In doing this, we take into account pre-merger mass loss and the halo switching problem (see the above-mentioned paper for a complete description). We also track various other properties of the galaxies through their histories, including star formation rate, black hole accretion rate, effective radius, colour, and metallicity, among many others. 

\section{Results}

\begin{figure*}
\begin{center}
\includegraphics[width=2\columnwidth]{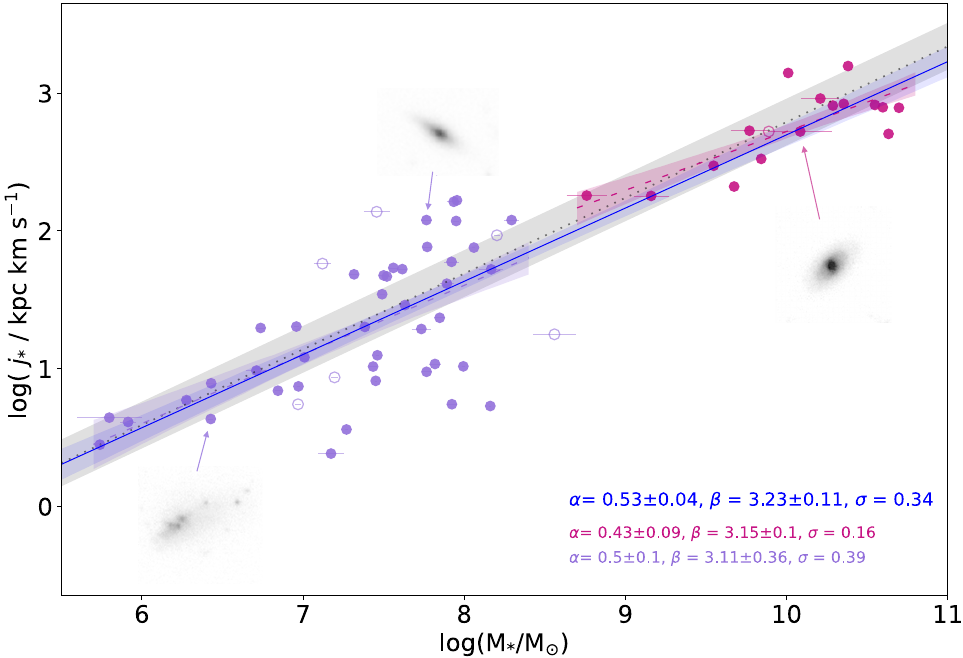}
   \caption{The $j_{*}$--$M_{*}$ relation for our observational sample, showing the SAMI control galaxies (magenta) and the dwarf galaxies (purple). The dark blue line and associated shaded region display the fit of the entire sample to Equation~\ref{logj-logM} along with its 68 percent credible interval. Fits for above and below $10^{8.5}\,\rm M_{*} (M_{\odot})$ (corresponding to the control and dwarf galaxy samples) are shown by the dashed purple and magenta lines respectively. The corresponding values and intrinsic orthogonal scatter ($\sigma$) for all three fits are given in the bottom right of the figure. The black dotted line and associated grey shaded region display the relation found by \citet{2018A&A...612L...6P}. Non-converging galaxies ($j_{*\rm tot}<0.8j_{\infty}$) are shown with an open circle. Example HSC $r$-band images of a large disk galaxy, a small disk galaxy, and an irregular dwarf galaxy are shown. The control galaxies closely follow the relation found by \citet{2018A&A...612L...6P} and our dwarf sample extends this relation down to below 10$^{6}\,\rm M_{*} (M_{\odot})$. }
 \label{j_m_obs}
\end{center}
\end{figure*}

\begin{figure}
\includegraphics[width=1\columnwidth]{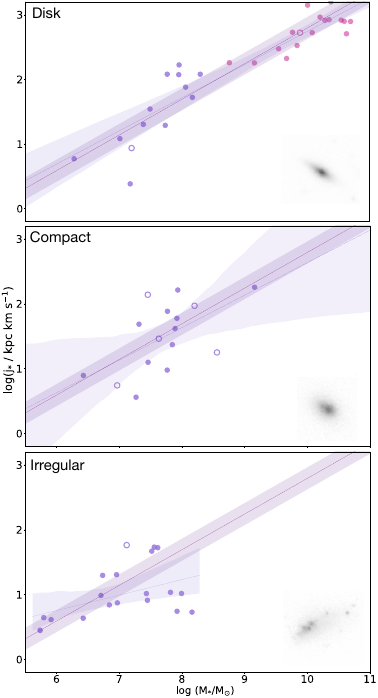}
   \caption{The $j_{*}$--$M_{*}$ relation with galaxies divided by their visually classified morphology, illustrating how the deviation from the previous relationship is morphology-dependent. The top panel shows disk galaxies, the middle panel shows compact galaxies and the lower panel shows clumpy and irregular galaxies.}
 \label{morphology}
\end{figure}

\begin{figure*}
\begin{center}
\includegraphics[width=2\columnwidth]{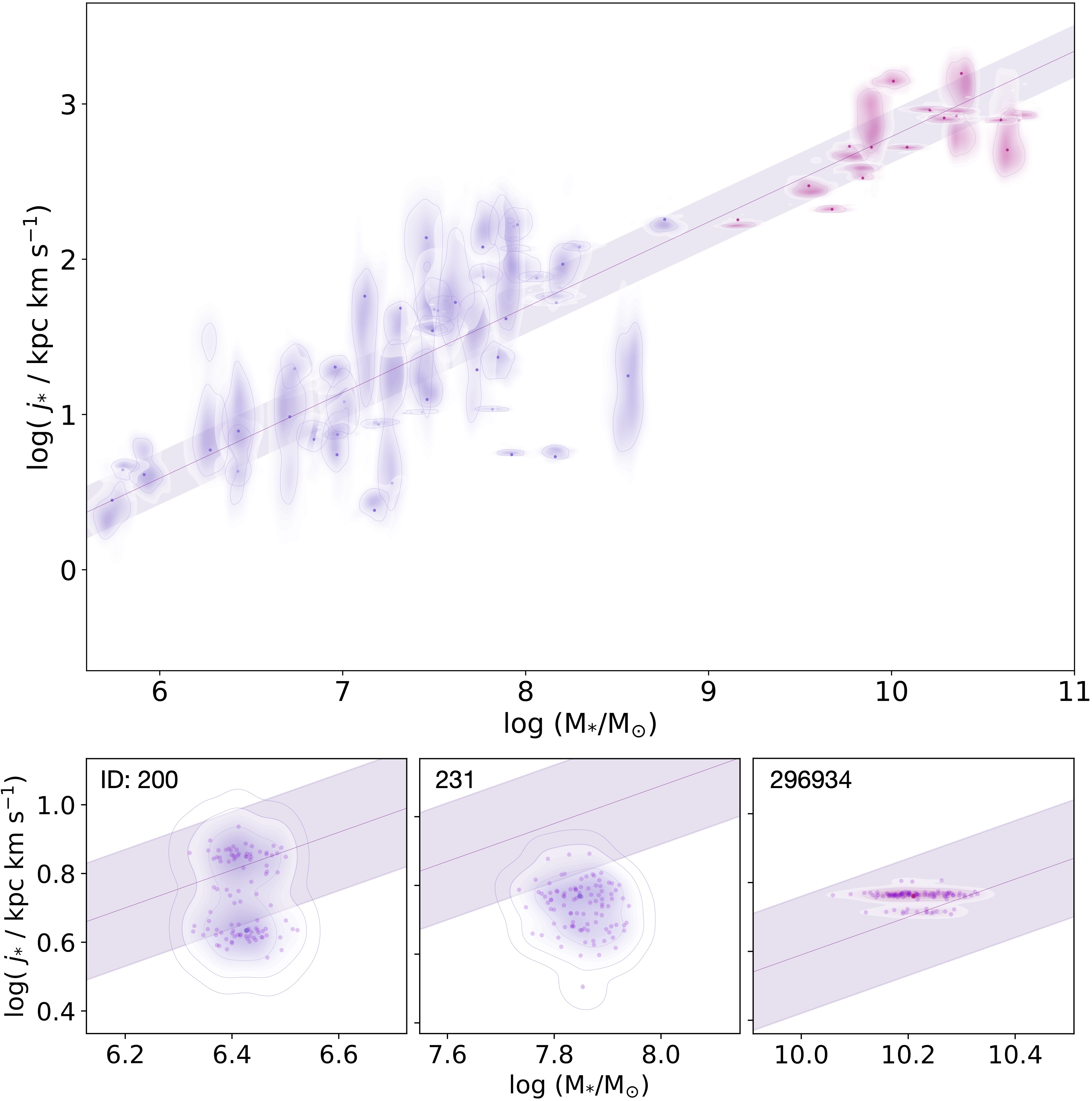}
   \caption{Distributions in $j_{*}$ after 100 Monte Carlo simulations for each galaxy. The top panel shows the MC distributions of each galaxy plotted together on the $j_{*}$--$M_{*}$ relation, with the colours as before; contours show 1 and 2 $\sigma$ from the centre of each distribution. The lower panels show example galaxies: left - a galaxy where the MC simulations led to a bimodal distribution in $j_{*}$; middle - a more typical dwarf galaxy; and right - a typical high-mass disk galaxy.}
 \label{MCfig}
\end{center}
\end{figure*}

\subsection{Fall relation for dwarf galaxies}

\begin{figure*}
\begin{center}
\includegraphics[width=2\columnwidth]{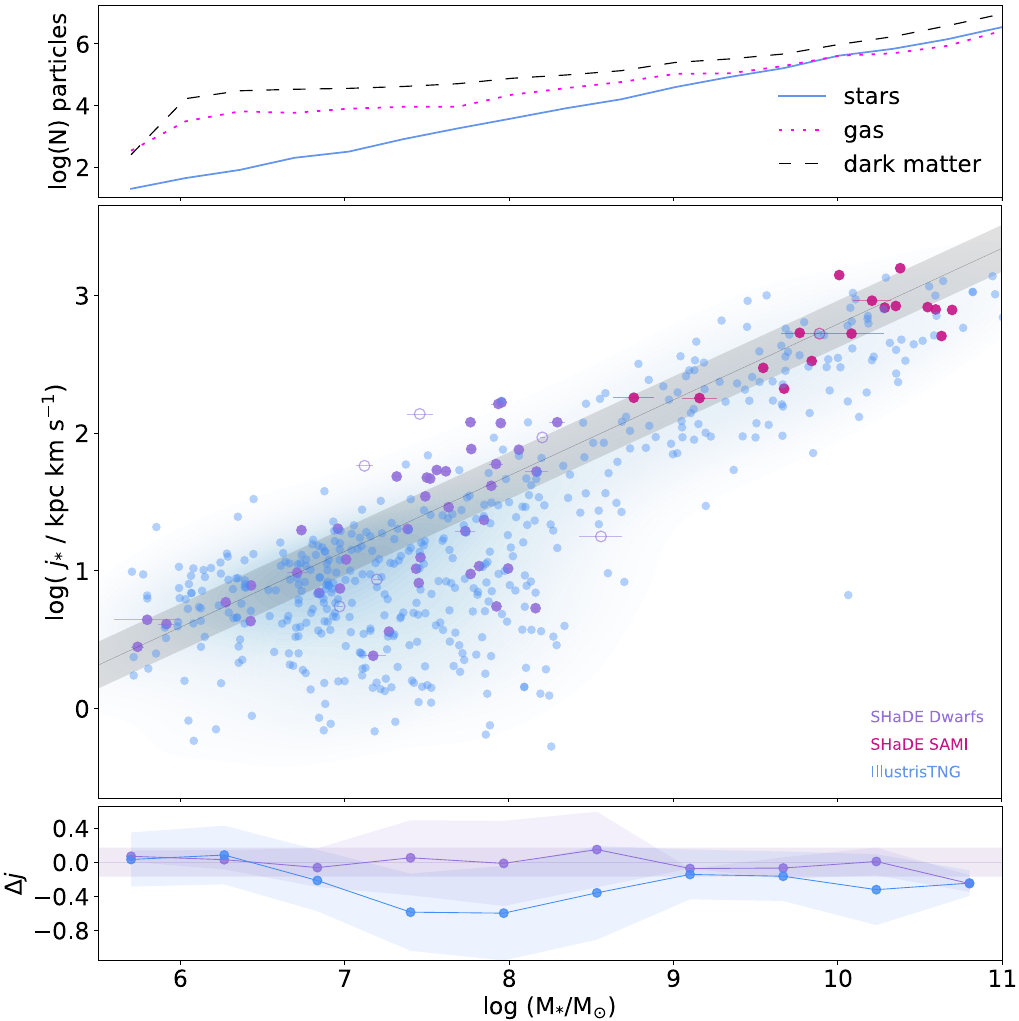}
   \caption{Comparing the $j_{*}$--$M_{*}$ relation for our observational sample with that from the IllustrisTNG sample. The central panel shows the same observational data points as Figure~\ref{j_m_obs}, with the simulation sample added as the blue points. The distribution of the simulated galaxies closely matches the observed galaxies, including the agreement with \citet{2018A&A...612L...6P} at higher masses and the extension to lower values of $j_{*}$ at lower masses. However the simulated galaxies have a greater range of $j_{*}$ values than the observed sample, with the lower-mass dwarf galaxies having an increased scatter towards lower $j_{*}$. This is highlighted in the bottom panel, which shows the median difference between $j_{*}$ and the expected relation from \citet{2018A&A...612L...6P} within mass bins centred on each point. The shaded regions show the standard error about each point. As an indication of how well resolved the simulated galaxies are, particularly for the low-mass end, the top panel displays the average particle number for each galaxy subhalo as a function of mass. The number of stellar particles decreases below 100 at around 10$^{6.2}\,\rm M_{*} (M_{\odot})$ while the number of gas particles is more than 1000 over the whole observed mass range.}
 \label{j_m_mod}
\end{center}
\end{figure*}

Figure~\ref{j_m_obs} shows the specific angular momentum for the SH$\alpha$DE galaxies, including both the high-mass `control' galaxies in red and the dwarf galaxies in blue. We compare these findings with the Fall relation derived from disk galaxies by \citet{2018A&A...612L...6P}, with $\alpha = 0.55$ and $\beta = 0.34$. The high-mass disk galaxies neatly follow the expected relation with little scatter. This illustrates that deriving $j_{*}$ from these H$\alpha$ emission line observations is able to recover $j_{*}$ values consistent with those of previous studies of higher-mass galaxies, and in addition, shows that the angular momentum of star-forming regions is tracing the angular momentum of the general stellar population. The dwarf galaxies in our sample, while featuring a larger degree of scatter particularly for stellar masses between $10^{7}$ M$_{\odot} < M_{*} < 10^{8}$ M$_{\odot}$, follow the relation of high-mass disk galaxies extrapolated to lower masses. 

Following \citet{2018A&A...612L...6P}, we carry out a log likelihood fit to Equation~\ref{logj-logM} accounting for the uncertainties in $j_{*}$ (as determined from Monte Carlo simulations, see Section~\ref{MC}) and intrinsic scatter orthogonal to the relation. We then employ the python package \texttt{emcee} \citep{2013ascl.soft03002F} and use a Monte Carlo Markov chain to sample the posterior distribution in order to determine the posterior 1$\sigma$ range. We find a best fit over the complete mass range of $\alpha = 0.53 \pm 0.04$ and $\beta = 3.23 \pm 0.11$, with an intrinsic scatter of 0.35\,dex (where the uncertainties correspond to the 68 percent credible interval). This is in close agreement with \citet{2018A&A...612L...6P} and shows that this relation can be extended down to stellar masses as low as $10^{6}\,\rm M_{*} (M_{\odot})$, thus covering 5 orders of magnitude in stellar mass.

Given that the maximum extent of our measurements varies across our sample, we restricted the radius to which $j_{*}$ is calculated to 1, 2 and 3 effective radii in order to determine how the varying radial extent may influence our results. While the absolute $j_{*}$ values across the whole sample increase as the radial limit increases and we include more of the outskirts, the observed result of a continuation in the $j_{*}$--$M_{*}$ relation from high-mass to low-mass galaxies remains consistent, showing that the varying radial extent of our data does not significantly impact our conclusions. 

Since our sample is selected to be star-forming, quenched dwarf galaxies may deviate or be offset from this relation (as is the case for the high-mass elliptical galaxies). As a test of whether there may be a significant difference between the populations, we look for a correlation within our sample between the star formation rate and the residuals from the fitted relation. We do indeed observe such a correlation, with dwarf galaxies above the relation tending to have higher star formation rates relative to those below (Spearman's correlation coefficient = 0.48, p-value = 0.001). Such a relationship is consistent with previous findings of a correlation between the neutral gas fraction and $j$ \citep{2022MNRAS.516.4043H}, with gas-rich galaxies having higher $j$ values. Separating our sample into two populations above and below the median specific star formation rate log(sSFR / yr$^{-1}$) = -9.45 and carrying out a log likelihood fit on each population, we find $\alpha = 0.56 \pm 0.07$, $\beta = 3.46 \pm 0.25$ for the population with higher star formation rates and $\alpha = 0.59 \pm 0.06$ and $\beta = 3.22 \pm 0.17$ for the population with lower star formation rates. While the normalisation of the relation is shifted to higher values for galaxies with higher star formation rates, the slopes are consistent with each other and both are consistent with the complete sample, within uncertainties. This suggests that our sample selection biases the normalisation but not necessarily the slope, however a quenched dwarf galaxy sample is needed to fully test this finding. We therefore emphasise that our results apply primarily to the star-forming dwarf galaxies, with additional observations being needed to study the quenched population.

Figure~\ref{morphology} shows the sample categorised by visual morphology; those that resembled disks are shown in Figure~\ref{morphology}a, compact galaxies in Figure~\ref{morphology}b, and clumpy irregular galaxies in Figure~\ref{morphology}c. Disk galaxies follow the global $j_{*}$--$M_{*}$ relation with a smaller degree of scatter (median offset = 0.035, $\sigma$ = 0.30) relative to the compact (median offset = 0.002, $\sigma$ = 0.42) and irregular (median offset = 0.003, $\sigma$ = 0.43) galaxies. Restricting to disk galaxies below $10^{8.5}\,\rm M_{*} (M_{\odot})$, the median offset becomes 0.093 with $\sigma$ = 0.35. Together, these results show that despite the more disordered structure of irregular galaxies and the lack of a clear disk in compact galaxies, these galaxy types continue to follow the $j_{*}$--$M_{*}$ relation to below $10^{6}\,\rm M_{*} (M_{\odot})$, albeit with increased scatter.

\subsection{Monte Carlo model fitting}
\label{MC}

To better characterise the uncertainties in our calculations of $j_{*}$, we performed a Monte Carlo simulation of our observations. For each galaxy, we created 100 random realisations where the velocity maps, mass and redshift were all randomly altered based on the measurement uncertainties in each respective property. For the velocity maps, the velocity value of each individual spaxel was replaced by a random number from a normal distribution centred on the observed value and with standard deviation equal to the uncertainty in the measurement for that spaxel. The redshift and mass of the galaxy was likewise replaced by a random number taken from a normal distribution centred on the original measurement and with the appropriate standard deviation. 

The resulting 100 realisations were then analysed in the same way as the original observations, including the fitting of a disk model and the subsequent calculation of $j_{*}$ from the deprojected maps. The distributions of $j_{*}$ from this process are displayed in Figure~\ref{MCfig}. There is great variety in the shapes and sizes of the resulting distributions. Many highly disk-like galaxies have a very small range of $j_{*}$ values, showing that in these cases the resulting disk model is very consistent between different realisations, and hence that it is robust. In other cases there is a large range of $j_{*}$ values, often for more irregular galaxies, showing that there is greater uncertainty in the modelling of these galaxies and reflecting the fact that it is difficult to model them with a disk. There are also cases where there are two favoured solutions in the model fitting, creating two regions in the $j_{*}$--$M_{*}$ plane; an example of this is shown in Figure~\ref{MCfig}~b). In this case the higher-$j_{*}$ realisations occur when the galaxy is modelled as being closer to face-on, meaning that its intrinsic rotational velocity has to be higher in order to match the value observed along the line-of-sight.  


\subsection{Testing the thin-disk assumption}
\label{thin-disk test}

As dwarf galaxies often have an irregular or thick-disk morphology, the method of deprojecting a thin kinematic disk, as used here, may not lead to realistic modelling of their intrinsic kinematics. We therefore test an alternative method for deprojecting the line-of-sight velocities for each dwarf galaxy based on the optically observed ellipticity (derived from multi-Gaussian fitting to SDSS imaging, see \citet{2020MNRAS.498.5885B}) and systematically varying the assumed intrinsic disk thickness ($q_0$) from 0 to 0.35. For higher $q_0$ values, our sample size becomes too small to carry out a meaningful fit due to the loss of galaxies (such as nearly edge-on disks) which cannot physically have a $q_0$ value at the set quantity. Using these deprojections, we then calculated $j_{*}$ as before. For low values of $q_0$, we found close agreement between the two methods (for $q_0$=0.05: median offset = -0.044\,dex, $\sigma$ = 0.062\,dex), showing consistency between the optically-derived and kinematically-derived disk deprojections. For each value of $q_0$ we fitted Equation~\ref{logj-logM} for $\alpha$ and $\beta$ to the resulting $j_{*}$--$M_{*}$ plane using the same method as before, and compared it to our original result for the dwarf galaxy population. We found these relations with varying $q_0$ to be consistent with our original result for the dwarf galaxies (M$_{*} \leq 10^{8.5} \rm \,M_{\odot}$) within uncertainties. This suggests that for the sample presented here, our derived slope remains consistent when varying the assumed disk thickness up to $q_0$ = 0.35.

\begin{figure*}
\begin{center}
\includegraphics[width=2\columnwidth]{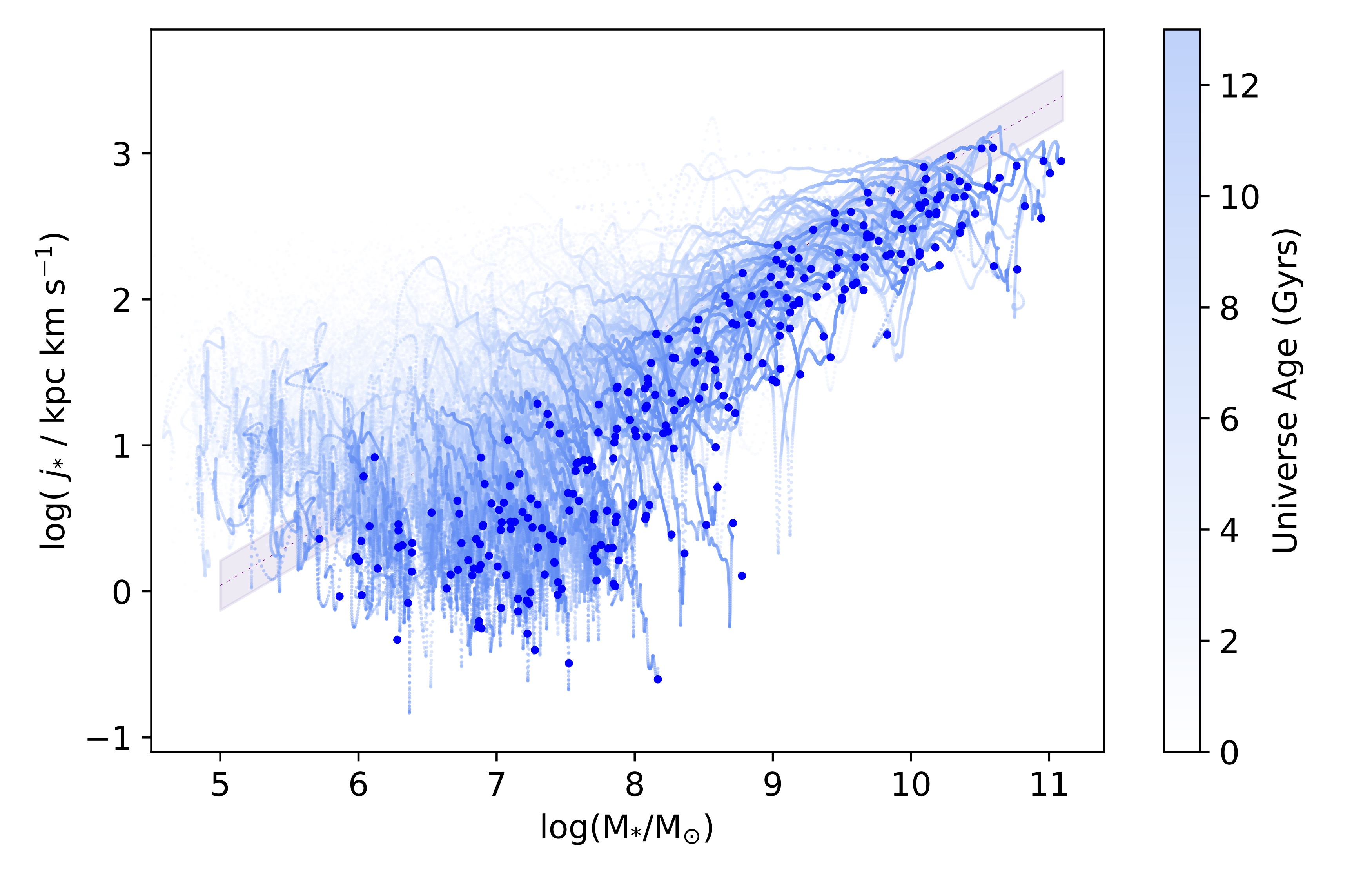}
   \caption{The $j_{*}$--$M_{*}$ relation for IllustrisTNG galaxies over their evolutionary history. To reduce clutter, we only show a random sub-sample. The mass and $j_{*}$ of each galaxy at every snapshot in the simulation is shown by a point, coloured by the time of that snapshot. The final positions of each galaxy are shown by the blue points. Galaxies initially forming with low masses gradually increase in mass and lose angular momentum; then, while some continue to lose momentum, others experience an increase in momentum as they continue to increase in mass. }
 \label{sim_evol_1}
\end{center}
\end{figure*}

\subsection{Comparison with SAMI and the effect of the PSF}

Since the 20 control galaxies in our sample were taken from the SAMI survey, we can also use the SAMI velocity maps as an observationally independent way to measure their angular momentum. The SAMI survey derived velocity maps from both their stellar component (using the stellar absorption lines) and from their ionised gas component (measured from the gas emission lines). With H$\alpha$ emission being a component of the gas emission, we expect the gas velocity field to more closely match what was observed in SH$\alpha$DE. We found that the angular momentum measured from the SAMI velocity maps was consistently lower than SH$\alpha$DE for both the stellar and gas component, and with the SAMI stellar component being lower than the SAMI gas component in most cases. To test whether this may be due to the different field of views of the two instruments, we overlaid the Argus footprint onto the SAMI maps, with its position and orientation matched to the SH$\alpha$DE observation. The resulting $j_{*}$--$M_{*}$ relation was tighter and more consistent with the SH$\alpha$DE relation, however the downward offset remained. This is likely due to the SAMI survey having a larger PSF (median full width at half-maximum (FWHM) of 2.04 arcseconds, \citet{2021MNRAS.505..991C}) relative to the SH$\alpha$DE survey (median FWHM of 0.88 arcseconds). The larger PSF acts to blur the velocity field, reducing the velocity gradient across the field and therefore reducing its apparent rotation (while increasing the apparent velocity dispersion).

To determine how the PSF of our observations may influence the derived $j_{*}$--$M_{*}$ relation, we tested for a correlation in the scatter of the $j_{*}$--$M_{*}$ relation with both the number of resolved elements in each galaxy and with their effective radii as measured in arcseconds on the sky. In relation to the number of resolved elements, we found no significant trend with the scatter from the relation (Spearman's correlation $p$-value = 0.48). We do however find a weak but significant trend between the scatter and effective radius in arcseconds (Spearman's correlation coefficient = $-$0.11, $p$-value = 0.026). This is in the opposite sense to what one may expect from PSF blurring, and may reflect the increased reliance on model extrapolation to 3 effective radii.

\begin{figure*}
\begin{center}
\includegraphics[width=2\columnwidth]{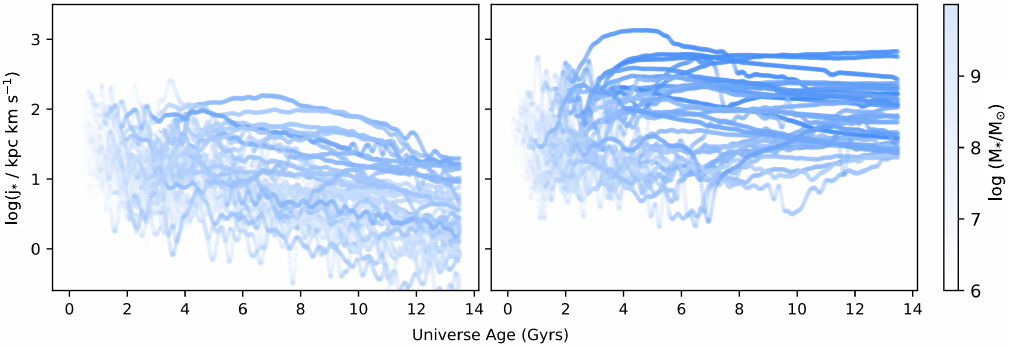}
   \caption{Evolution of $j_{*}$ as a function of time for all galaxies, divided into those with a final mass below $10^{8}\,\rm M_{*} (M_{\odot})$ on the left and above $10^{9}\,\rm M_{*} (M_{\odot})$ on the right. The low-$j_{*}$ galaxies, after an initial rapid decrease in $j_{*}$, continue to lose angular momentum at a steady rate, with high fluctuations in $j_{*}$ for each individual galaxy. The high-$j_{*}$ galaxies, meanwhile, go through the same initial loss of angular momentum before experiencing a rapid increase, eventually reaching a steady higher-$j_{*}$ state. }
 \label{sim_evol_2}
\end{center}
\end{figure*}

\subsection{Model Fall relation}

We then attempted to reproduce the Fall relation using our sample of galaxies from the IllustrisTNG-50 simulation. To be consistent with our observational sample, $j_{*}$ was calculated from the velocity maps made to resemble the SH$\alpha$DE observations (deriving the maps from the H$\alpha$ emitting gas); the resulting relation is shown in Figure~\ref{j_m_mod}, along with the SH$\alpha$DE observations from Figure~\ref{j_m_obs}. Above a stellar mass of  $10^{9}\,\rm M_{*} (M_{\odot})$, the simulated galaxies closely follow the relation found by \citet{2018A&A...612L...6P}, and are in good agreement with the high-mass SH$\alpha$DE control sample. This agreement of IllustrisTNG with previous observational studies in this mass range has been seen previously by \citet{2015ApJ...804L..40G} (who used the original Illustris simulation), and  \citet{2022MNRAS.512.5978R},  \citet{2022ApJ...937L..18D} and \citet{2023MNRAS.526..808H} (all using IllustrisTNG). However, all of these studies were restricted to masses above $10^{9}\,\rm M_{*} (M_{\odot})$. Below this mass there is an increased scatter towards lower $j_{*}$ values, with the upper edge continuing to follow the observed relation. Alternatively this could be seen as a drop and flattening of the relation, as reflected in the residual plot in the bottom panel of Figure~\ref{j_m_mod}. The low-$j_{*}$ dwarf galaxies may be missing from our observational sample if they are faint or have low surface brightness, or if they reside in higher-density environments not probed by our nearby dwarf galaxy sample. 

We compared our derived $j_{*}$ values to the intrinsic values of the galaxies as calculated from their particles directly. We found relatively close agreement in $j_{*}$ between our observation-equivalent values and the star-forming particles, with the residuals (observed\,$-$\,intrinsic) having a median bias of 0.08 and a scatter (one standard deviation) of 0.39. We found our observed values to overestimate the intrinsic stellar $j$ (median bias = 0.44, scatter = 0.43), highlighting that the star-forming component has elevated angular momentum relative to the broader stellar component. We also find the relation between our observed values and the intrinsic $j_{*}$ to deviate from a one-to-one relation, finding a gradient $m$ of $0.80\pm0.02$ for $j_{observed}$\,=\,$mj_{intrinsic}$, largely driven by an increase in scatter in the intrinsic stellar $j$ at low mass. This suggests the difference in $j$ between the star-forming component and stellar component may be greater for dwarf galaxies relative to high-mass galaxies, though again we stress the potential resolution limitations of the simulation at lower masses.

In the process of calculating the intrinsic angular momentum, we found the total angular momentum vector of all stellar particles within each galaxy, allowing us to determine the inclination angle as projected onto the observed $x$--$y$ plane. We found no significant correlation between the inclination angle and the residuals of the observed\,$-$\,intrinsic $j$ (Spearman's coefficient = 0.01, $p$-value = 0.76), implying minimal bias in our observed values due to inclination. As a kinematic measure of the disk thickness, we calculated a simplified intrinsic $v/\sigma$ value (where $\sigma$ is the velocity dispersion) by taking the ratio of the sum of particle velocity components within the plane of rotation to the sum of velocity components outside this plane, and found there to be no significant correlation with the residuals (Spearman's coefficient = $-$0.02, $p$-value = 0.62) We also find no significant correlation with effective radius (Spearman's coefficient = $-$0.06, $p$-value = 0.24). However as stated through this paper, when applying these findings to our observational results, potential resolution effects within the simulation should be kept in mind.

The decrease in $j_{*}$ for low-mass galaxies, and the apparent flattening of the relation below $10^{8}\,\rm M_{*} (M_{\odot})$ may also be a result of the simulation's limited resolution. Galaxies with stellar masses around $10^{8}\,\rm M_{*} (M_{\odot})$ consist of a few thousand stellar particles, while $10^{7}\,\rm M_{*} (M_{\odot})$ consist only of a few hundred stellar particles, meaning the galaxies' structures are less resolved which may lead to the loss of coherent motions. Similarly in the mass-radius relation, as discussed above, there is a flattening below $10^{8}\,\rm M_{*} (M_{\odot})$. As a test of whether the observed low $j_{*}$ values among the dwarf galaxies are due to resolution effects, we carried out the same analysis using the lower-resolution TNG100 and TNG300 simulations. While an increasing degree of scatter is seen as we go from TNG50 to TNG300, the distribution of galaxies on the $j_{*}$--$M_{*}$ plot remains consistent. In particular, we do not see a change in the mass at which galaxies begin to scatter below the expected relationship---if this was purely a resolution effect, we may expect to see the downward scatter in $j_{*}$ begin at higher galaxy masses as the simulation's resolution decreases. This suggests that the measured $j_{*}$ seen among the dwarfs may be due to physical processes within the simulation rather than the limited resolution. We also note that the H$\alpha$ kinematic maps for dwarf galaxies around $10^{6}\,\rm M_{*} (M_{\odot})$ remain physical, with coherent rotation clearly visible in well-ordered systems.

\begin{figure*}
\begin{center}
\includegraphics[width=1.9\columnwidth]{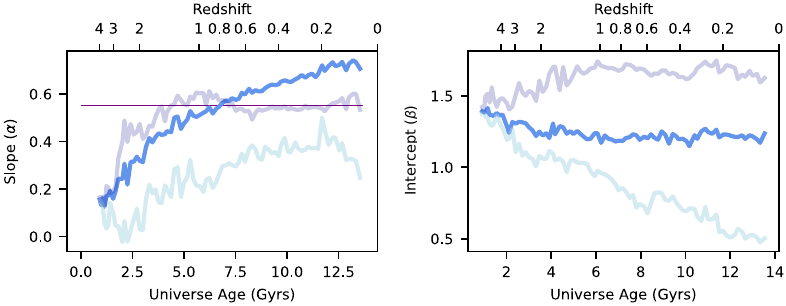}
   \caption{Evolution of the $j_{*}$--$M_{*}$ relation with the age of the Universe. The left panel shows the slope of the relation while the right panel shows the intercept. Light blue lines show the evolution of galaxies which have a final mass below $10^{9}\,\rm M_{*} (M_{\odot})$, the dark blue lines shows the evolution of those which masses above $10^{9}\,\rm M_{*} (M_{\odot})$ and the purple line shows the evolution for the complete sample. The overall slope is shallower at high redshifts and increases with time until it reaches the present-day value at a redshift of around 1.5, and becomes stable by $z$ = 0.6. The slope amongst the high-mass galaxies is seen to continually increase with time, while the slope amongst the dwarfs increases at a slower rate before decreasing towards the present day.}
 \label{evoln_j-M}
\end{center}
\end{figure*}

\subsection{Evolution of \texorpdfstring{$j_{*}$}{j}}
\label{sec_j_evolution}

Figure~\ref{sim_evol_1} displays the evolution of the specific angular momentum for a sample of galaxies in our simulation sample. The evolution is plotted in the $j_{*}$--$M_{*}$ plane, showing the stellar mass and specific angular momentum at each simulation snapshot for every galaxy. Considering the computational burden of generating mock observations, here we use the true $j_{*}$ calculated from the individual particles in each galaxy. The particles used here are the gas particles weighted by their star formation rate, in order to match the H$\alpha$ component measured in our observation sample. The combined evolution of all galaxies forms a flat distribution at the low-mass end, a large dip to low angular momentum around $10^{7} - 10^{8.5}\,\rm M_{*} (M_{\odot})$ and an upward slope at higher masses. Galaxies initially form with low masses and with a relatively significant and highly fluctuating amount of angular momentum, creating the broad low-mass arm of the plot. As they grow in mass, most galaxies either fluctuate around a relatively constant value or lose angular momentum, creating the downward dip with increasing mass. At around $10^{7} - 10^{8}\,\rm M_{*} (M_{\odot})$ there is a transition to a rising relation between $j_{*}$ and M$_{*}$, where the galaxies appear to be gaining angular momentum as their mass further increases. This creates the upward-sloping right-hand side of the figure. From this pattern, it appears that galaxies first lose angular momentum and then either maintain or begin gaining angular momentum once they grow past a certain mass range. We note that here we are only considering star-forming galaxies, leaving out quenched elliptical galaxies which likely have differing histories. In contrast, lower-mass galaxies appear to always be losing angular momentum while gaining mass and never reach this transition point. Finally, a few galaxies are seen to dip down in $j_{*}$ and to move to the left, suggesting that they are being stripped of their outer rotating disk. 

Evidently, despite the apparent simplicity and tightness of the Fall relation as observed in this and various other studies in the present day, particularly for higher masses, the evolution of $j_{*}$ leading up to the present-day values is very complex and dynamic. We emphasise here that we are looking at galaxies which are selected as star-forming disk galaxies at the present day; high-mass galaxies which undergo more disruptive mergers and evolve into dispersion-dominated ellipticals would experience different evolutionary histories in $j_{*}$ and fall below the $j_{*}$--$M_{*}$ relation at the present day. 

Figure~\ref{sim_evol_2} displays the same information as in Figure~\ref{sim_evol_1}, but instead we are plotting $j_{*}$ as a function of time for each galaxy. The galaxy sample is divided into those which have a final mass below $10^{8}\,\rm M_{*} (M_{\odot})$ and above $10^{9}\,\rm M_{*} (M_{\odot})$. This more clearly illustrates that there appears to be two main evolutionary paths for the angular momentum of these galaxies, though we note that there is a continuum between these paths rather than a clear separation.

\begin{figure*}
\begin{center}
\includegraphics[width=2\columnwidth]{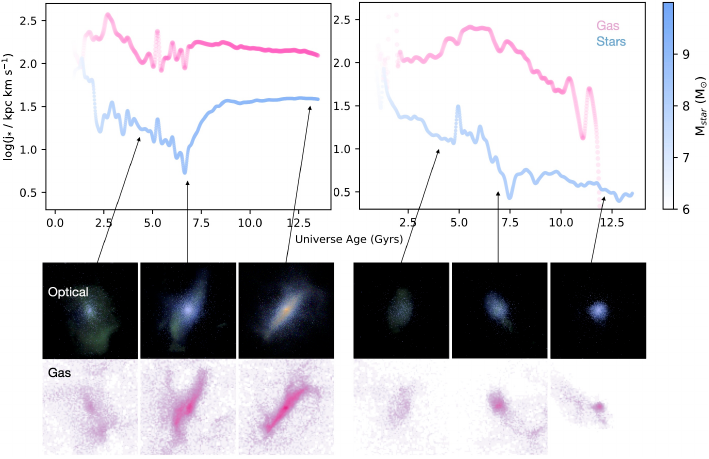}
   \caption{Examples of a galaxy with a low final value of $j_{*}$ (right) and with a high final value (left). The top panels show the evolution of $j_{*}$ with time, with the blue sequence showing the stellar component and the pink sequence showing the gas component. The bottom two rows show the optical image as modelled by SKIRT (top row) and the distribution of the gas component (bottom row) at three different times.}
 \label{evol_ex}
\end{center}
\end{figure*}

Galaxies with final masses below $10^{8}\,\rm M_{*} (M_{\odot})$ initially form with higher $j_{*}$, then experience a steady loss of $j_{*}$ with the passage of time. Galaxies with final masses above $10^{9}\,\rm M_{*} (M_{\odot})$ also experience an initial loss of angular momentum, then in many cases undergo a rapid transition or jump to a state of higher angular momentum. The fluctuations in $j_{*}$ also decrease significantly after this transition, and once it is completed the galaxies are able to maintain a very steady value of $j_{*}$, resulting in a relatively constant mean and dispersion in $j_{*}$ through time for this group. Most of these galaxies have entered a steady state by 6\,Gyrs after the Big Bang (corresponding to a redshift of 1), which is consistent with the lack of significant evolution in the $j_{*}$--$M_{*}$ relation in observations over this redshift range for high-mass galaxies. The jump up to higher $j_{*}$ does not occur at a particular critical mass; rather, it occurs once the galaxy has grown into the stellar mass range between $10^{8}$ and $10^{9}\,\rm M_{*} (M_{\odot})$. Galaxies that never reach this mass never experience this transition, and so continue to lose angular momentum. However, as time goes on galaxies of lower and lower mass experience this jump in $j_{*}$, and so perhaps some lower-mass galaxies haven't yet had enough time to experience this transition. 

It remains possible that this transition in $j_{*}$ corresponds to a resolution effect, for example the crossing of a threshold in the number of particles within the galaxy resulting in an alteration of internal dynamics. As a test on whether the jump in $j_{*}$ is a result of resolution limitations, we trace the evolution of $j_{*}$ for high-mass galaxies within the lower-resolution TNG100 simulation and find similar behaviour, with a transition to a high-$j_{*}$ disk-dominated galaxy occurring from $10^{8}$--$10^{9}\,\rm M_{*} (M_{\odot})$; an example from TNG100 is provided in Figure~\ref{resolution} in the Appendix. This consistency in the $j_{*}$ transition between TNG50 and TNG100 suggests that this may not be due to resolution limitations. We note also that within TNG50, the timing of this transition does not correspond to galaxies reaching a particular particle number as they grow in mass.

To determine how the Fall relation evolves over time within our simulation sample, we performed a fit of Equation~\ref{logj-logM} for each snapshot over the complete sample and for galaxies above and below $10^{9}\,\rm M_{*} (M_{\odot})$. The evolution of $\alpha$ and $\beta$ over time is displayed in Figure~\ref{evoln_j-M}. We find that despite the varied behaviour of individual galaxies, the value of $\alpha$ over the full sample reaches the present-day value at around $z$ = 1.5 and becomes stable by $z$ = 0.6. 

\subsubsection{Transition in $j_{*}$ and disk formation}

To illustrate the dynamical changes relating to the increase in $j_{*}$ we compare individual galaxies which do or do not experience this transition. Figure~\ref{evol_ex} shows two such galaxies. The first example (on the left side of the figure) is a disk galaxy with a final stellar mass of $10^{8.8}\,\rm M_{*} (M_{\odot})$ and the second is a dwarf galaxy with a final mass of $10^{7}\,\rm M_{*} (M_{\odot})$. These two galaxies initially follow similar behaviour in their evolution of angular momentum, starting with a relatively high $j_{*}$ which gradually decreases with time. The stellar and gas distributions show that at this early stage both galaxies are irregular, disordered systems with no signs of any disk components, with the gas component appearing clumpy and turbulent. However, the $j_{*}$ evolution of the two galaxies diverge at around 6.2\,Gyrs, when the higher-mass galaxy begins experiencing a rapid increase. This is associated with the gas beginning to collapse into a plane of common rotation, which appears nearly edge-on in this orientation. Once the galaxy settles into a new dynamical state with ${\rm log}(j_{*}) \approx 1.55$, the gas is now in a tight disk, and with new stars now forming within this disk, a disk component is also clearly evident in the stars. In the case of the lower-mass dwarf galaxy, as $j_{*}$ continues to decrease, the gas remains turbulent and the stellar distribution remains irregular or elliptical with no signs of a developing disk. 

\begin{figure}
\includegraphics[width=1\columnwidth]{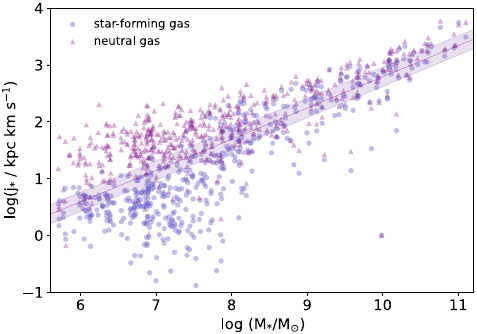}
   \caption{Comparison of the intrinsic $j$ values for the neutral gas (purple triangles) and stellar (blue circles) components. The $j$ value shown here is calculated directly from the individual gas and star particles respectively. In the case of the neutral gas component, $j$ was calculated from all gas particles weighted by their neutral gas fraction. The Fall relation, as derived by \citet{2018A&A...612L...6P}, is shown by the purple line and accompanying shaded region. Below $10^{8}\,\rm M_{*} (M_{\odot})$, the gas component has higher $j$ relative to the stellar component, while above this mass, the gas and stellar components share a common disk and have similar $j$.  }
 \label{neutral_j}
\end{figure}

\section{Discussion}

\subsection{Comparison with previous HI studies}

As mentioned in the introduction, most previous studies of the specific angular momentum within dwarf galaxies have been carried out using HI observations. Our observational results are consistent with \citet{2017MNRAS.472.4551E}, in that they found their dwarf galaxies to follow the same relation as that determined for the same component in high-mass disk galaxies (in their case HI). The main difference for our study, aside from measuring a different component and finding a consistent result, is the extension of this continuation down to galaxy masses of 10$^{5.7}\,\rm M_{*} (M_{\odot})$. 

However, like \citet{2017MNRAS.472.4551E}, our results are in tension with the earlier studies by \citet{2017ApJ...834L...4B}, \citet{2017MNRAS.467.3856C} and  \citet{2018MNRAS.479..228K} who found higher values of $j$ for the dwarf galaxy population. \citet{2017MNRAS.472.4551E} argue that this difference, at least in the case of \citet{2017ApJ...834L...4B}, is due to the extrapolation of the HI rotation curve to radii beyond that where the rotation curve is well constrained. \citet{2017MNRAS.467.3856C} included a more limited sample of five dwarf galaxies, which were additionally selected to be gas-rich dwarf galaxies with a HI to stellar mass ratio greater than eight. We note that, as found by \citet{2022MNRAS.516.4043H}, $j$ increases with HI fraction for a given mass, and so it may be possible that this selection led to a sample of dwarf galaxies with higher-than-average angular momentum. \citet{2018MNRAS.479..228K} calculated $j$ for 11 dwarf galaxies residing in the void. Such dwarfs, as they have noted, have also been found to have higher HI gas fractions, and so it is possible that they have selected dwarf galaxies which have a higher degree of angular momentum. They noted that their results are consistent with the non-void galaxies studied in \citet{2017ApJ...834L...4B} and \citet{2017MNRAS.467.3856C}; however, as noted above, the dwarfs in \citet{2017MNRAS.467.3856C} were selected for high HI fractions and \citet{2017ApJ...834L...4B} may have measured higher $j$ values for other reasons. 

To further investigate the compatibility of our results with previous HI studies, we looked to use our simulation sample to determine how the angular momentum of the HI and stellar components relate to each other. IllustrisTNG models the fraction of neutral gas within each gas cell, however the differentiation of neutral hydrogen into atomic and molecular states isn't directly modelled. A full treatment would require detailed post-processing of the simulation data (see for example \citealp{2018ApJS..238...33D}), which we leave for future consideration. As an initial indication of potential differences between our studies, here we make a simplification by considering the fact that the mass fraction of molecular to HI gas in dwarf galaxies is typically less than 10 per cent \citep{2014A&A...564A..67B} and measure $j$ for the entire neutral gas component. We do this by calculating $j$ for the gas particles weighted by their neutral gas fractions; the resulting $j$ values for this component, along with the $j$ values for all gas particles which are star forming, are displayed in Figure~\ref{neutral_j}. The two components have similar $j$ values for the higher-mass galaxies, however below 10$^{8} \,\rm M_{*} (M_{\odot})$ the neutral gas component is seen to have higher $j$ values relative to the star-forming gas. This is an indication that measuring $j$ from the neutral gas component may lead to higher measured values of $j$ relative to the stellar component. 

The difference between the gas and stellar components could be due to dynamical heating of stellar populations after formation, or by the formation of stars occurring within lower-angular-momentum gas which has collapsed further towards the centre of the galaxy. Alternatively, this may also be due to the low baryonic particle numbers in the simulation, leading to a disproportionate transfer of heat from the dark matter halo to the baryonic component. While the consistency between our results for TNG50, TNG100 and TNG300 suggest that this may not be a dominating factor, higher resolution simulations will be helpful in removing this uncertainty. 

\subsection{The cause of the jump in \texorpdfstring{$j_{*}$}{j}}

In this subsection we consider possible causes of the jump in j: resolution effects, galaxy mergers, bursty star formation and the accretion mode of gas. When following the evolution of $j_{*}$ back through time in the simulation, we found that while all galaxies initially lose angular momentum over time, some then experience a significant and sometimes very rapid rise to a higher angular momentum state. Once in this state, these galaxies then maintain a relatively constant $j_{*}$, while those which do not experience this jump continue to lose angular momentum. Distributions of the star and gas components show that this jump corresponds to the formation of a stable rotating disk. 

It may be possible that this is due to the resolution of the simulation, with stable disks appearing once the galaxy passes a certain mass threshold, and hence a certain particle number threshold in the simulation. However, we found that the jump in $j_{*}$ occurs over a wide range of particle numbers, suggesting that it isn't triggered by a certain particle number limit. The fact that the final $j_{*}$ values of our simulation sample, resulting from this divergent evolution, closely agree with our observational measurements also supports a physical nature for these different evolutions of $j_{*}$. 

Galaxy mergers can have a significant and rapid effect on a galaxy's angular momentum. In particular, gas-rich (`wet') mergers can significantly increase $j_{*}$ \citep{2018MNRAS.473.4956L}. \citet{2022MNRAS.517.3459C} also found that certain merging configurations can lead to a very sudden rise in $j_{*}$ followed by a stable period. However, while some of our galaxies do experience minor and major merger events, many of those which experience a rapid rise in $j_{*}$ do not show any signs of mergers. They are isolated galaxies at the centres of their own group halos, and their tracks through space-time show no dynamical motions caused by the pull of any nearby galaxy. Therefore mergers or interactions with neighbouring galaxies are not the cause of the transition in $j_{*}$ that we see. 

Bursty star formation has been suggested as being disruptive for disks, due to the short periods of intense star formation and associated winds and radiation \citep{2023MNRAS.519.2598G}. The formation of a disk may therefore correspond to a transition to a smoother, more continuous rate of star formation. An initial investigation into the star formation activity in our simulation galaxies showed no significant change in the variability of star formation during the time in which they transition to a state of higher angular momentum, though this should be investigated more carefully in future work. 

\citet{2023MNRAS.525.2241H} showed using the FIRE simulations that a change in the shape of the gravitational potential corresponds to the formation of a disk. More specifically, when the central matter concentration of the galaxy increases beyond a certain point, the circular velocity profile changes from a profile which initially increases with radius to one which is decreasing with radius. We therefore investigated if there was a correlation between the sudden increase in $j_{*}$ we see in IllustrisTNG and the shape of the total gravitational potential. While we did see a change in the circular velocity profile at the same time as the jump in some cases, we did not see such a correlation across the full sample of simulated galaxies. 

The source and nature of gas falling into the central regions of the galaxy also potentially plays an important role. The initial gas inflow is thought to be in the form of high-angular-momentum cold gas accretion \citep{2013ApJ...769...74S}, which is often not aligned with the angular momentum of the dark matter halo, leading to a decrease in the overall specific angular momentum of the baryons during this phase \citep{2017ASSL..430..249S,2022MNRAS.514.5056H}. The stellar population forming out of this gas would subsequently also have a low specific angular momentum. As the galaxy increases in mass, a hot, virialised gaseous halo forms around the galaxy, acquiring the angular momentum of the host halo. Gas then falls into the centre isotopically from this hot halo and cools, creating a more constant, uniform source of gas inflow with aligned angular momentum onto the inner galaxy \citep{2022MNRAS.514.5056H}. This gas carries the angular momentum of the outer halo into the central regions of the galaxy, forming a high-angular-momentum disk. 

The recent study by \citet{2025arXiv250800991B}, carried out using the FIREBox simulation (a cosmological simulation with a volume of 22.1 cMpc$^{3}$), identified a transition region between low-mass dispersion-dominated systems and high-mass disk-dominated systems within the stellar mass range of $10^{9}<M_{*}<10^{10}$ M$_{\odot}$. Within this region the fraction of stars on disk-like orbits increased, which was taken as an indication of disk formation. This transition corresponded to a transition from bursty to steady star formation, a deepening of the gravitational potential and the formation of a hot gaseous halo, all of the main factors discussed above. While \citet{2025arXiv250800991B} found the end of bursty star formation to most strongly correlate with the beginning of disk formation, all three factors likely play a role in the formation of a disk and an associated rise in $j_{*}$. Similarly, while we didn't find any of these processes to play a singular role in the transformation seen in this work, a combination of all three may work together to create the transition in $j_{*}$ and allow the formation of a disk.

Interestingly, \citet{2025arXiv250800991B} found the transition range to occur an order of magnitude above the range where we have observed an increase in $j_{*}$ in this work ($10^{8}<M_{*}<10^{9}\,M_{\odot}$. This could indicate that the transition in $j_{*}$ begins before disk formation becomes evident in other parameters; however a consistent comparison would need to be made between these simulations for these findings to be compared more directly. 

\subsection{Implications for the understanding of galaxy angular momentum}

Our finding that the low-mass star-forming dwarf galaxies follow the same relation as that of the high-mass disk galaxies suggests that the process of angular momentum build-up through tidal torques, and the subsequent transfer and conservation of this angular momentum to the baryonic component, continues to be independent of the galaxy mass down to stellar masses of 10$^{6}\,\rm M_{*} (M_{\odot})$. This suggests that the high dark matter fraction, shallow potential wells (and subsequently more significant impacts of feedback processes) does not significantly alter the transfer and conservation of angular momentum within the baryonic component. This continuation of the relation, as found for high-mass disk galaxies, is also in spite of the fact that the lowest mass dwarfs in our sample have disordered, irregular morphologies. This may not be expected, as for galaxies above 10$^{9}\,\rm M_{*} (M_{\odot})$, many studies have found significant offsets in the $j_{*}$--$M_{*}$ relation with morphology (as detailed in the Introduction), with elliptical and bulge-dominated galaxies having lost angular momentum relative to (and after their transformation from) disk-dominated galaxies. Despite their chaotic structure and turbulent kinematics, irregular galaxies evidently retain a significant amount of their halo's original angular momentum within their central star-forming components. 

In terms of the evolution of $j_{*}$ with time, previous studies have shown the subsequent evolution can have a significant impact on the angular momentum initially imparted on halos through tidal torques; in particular, merger events which disrupt the disk/s and transform it into an elliptical galaxy greatly reduce the coherent rotational motion of the stars and result in a significant decrease in $j_{*}$. It is evident from Section~\ref{sec_j_evolution} that high-mass disk galaxies also have a very varied and dynamic evolution in $j_{*}$, yet despite this complexity, star-forming galaxies which maintain their disk morphology fall into a tight $j_{*}$--$M_{*}$ relation by the present day. 

\section{Summary and Conclusion}

In this work we aimed to determine if the relationship between a galaxy's mass and specific angular momentum ($j_{*}$), known as the Fall relation, extends down to the low-mass dwarf galaxies. To achieve this, we studied 49 star-forming dwarf galaxies and 20 high-mass control galaxies observed by the SH$\alpha$DE survey and measured their specific angular momentum from H$\alpha$-derived velocity maps and continuum-derived mass distributions. We found that our high-mass control galaxies follow the same relation as found for \citet{2018A&A...612L...6P}, showing the consistency of our tracer with previous studies. We further found that our dwarf galaxy sample continues to follow this relation down to $10^{5.7}\,\rm M_{*} (M_{\odot})$, with an increase in scatter below $10^{8}\,\rm M_{*} (M_{\odot})$. For our complete sample, we find $\alpha = 0.53 \pm{0.04}$ and $\beta = 3.23 \pm{0.11}$. The continuity in the relation to low masses, at least in the case of star-forming dwarf galaxies, supports the idea that the buildup of angular momentum in galaxies is scale-free, with the high dark matter fraction and shallow gravitational potential wells not having a significant effect on the build-up and conservation of angular momentum in the baryonic component through the galaxy formation process. We find a correlation in the residual of the relation with star formation rate, with higher $j$ values being associated with higher star formation rates; by analogy with the high-mass galaxy population, quenched dwarf galaxies will likely be offset from the observed relation. 

We then followed up these findings with an investigation into dwarf galaxies in the IllustrisTNG-50 cosmological simulation. To make a direct comparison with our observational results, we used the star-forming gas particles in each galaxy to create flux and kinematic maps resembling the SH$\alpha$DE observations. We then applied the same methods to the simulated observations to determine an observationally-equivalent value of $j_{*}$. We found that while the simulated low-mass dwarfs also generally follow the extended Fall relation, with a flattening of the relation and increased scatter towards lower-$j_{*}$ values below $10^{9}\,\rm M_{*} (M_{\odot})$. This may be caused by the resolution limitations of the simulations, with lower-mass galaxies being resolved by too few particles. However, we note that we find consistency in this distribution when we perform the same analysis using the lower resolution TNG100 and TNG300 simulations. Within the simulation, we then confirm that the angular momentum of the star-forming gas follows that of the stellar component in general, and that the neutral gas as a whole has a higher value of $j$ relative to the stars. This latter result, along with previous observational studies including \citet{2023MNRAS.519.1098C} and the fact that H$\alpha$ is less extended relative to HI, explains why we find lower $j$ values using H$\alpha$ compared to previous HI studies of dwarf galaxy angular momentum. We then followed the simulated galaxies back through their evolutionary history. We found that while all galaxies initially lose angular momentum with time, some galaxies experience a sudden and often rapid increase in $j_{*}$ when they are within the stellar mass range of $10^{8}$--$10^{9}\,\rm M_{*} (M_{\odot})$ before settling into a steady state with a relatively constant value of $j_{*}$. This transition may also be due to resolution limitations of the simulation, though we note that the same transition, in the same mass range, is found in the lower-resolution TNG100 simulation. 

Assuming a physical origin, we investigated many potential mechanisms for this sudden change in $j_{*}$ experienced by the higher-mass galaxies, but were unable to conclusively determine a single cause of this change. The transition is likely caused by a combination of factors including a change in the burstiness of star formation, deepening gravitational wells and a change in the mode of gas accretion. In order to fully investigate this transition, high-resolution optical observations of dwarf galaxies are needed in the stellar mass range of $10^{8}$--$10^{9}\,\rm M_{*} (M_{\odot})$, along with resolved HI observations allowing for the velocity mapping of the surrounding HI gas. Such observations would provide great insight into what causes a galaxy to undergo a state change in specific angular momentum, and potentially reveal the conditions necessary for the formation of disks in galaxies. No matter what the cause may be, the apparent simplicity and tightness of the Fall relation in the present day, observed here over 5 orders of magnitude in stellar mass, belies a complex and dynamic evolution of $j_{*}$. 

\section{Acknowledgements}

SD and SMS acknowledges funding from the Australian Research Council (DE220100003). LC acknowledges support from the Australian Research Council Discovery Project funding scheme (DP210100337). Parts of this research were conducted by the Australian Research Council Centre of Excellence for All Sky Astrophysics in 3 Dimensions (ASTRO 3D), through project number CE170100013. We would like to thank Aaron Romanowsky and Pavel E. Mancera Piña for helpful discussions on this paper. We thank the reviewer for helpful feedback which improved the paper. 

\bibliographystyle{mnras}


\appendix

\begin{figure*}
\includegraphics[width=2\columnwidth]{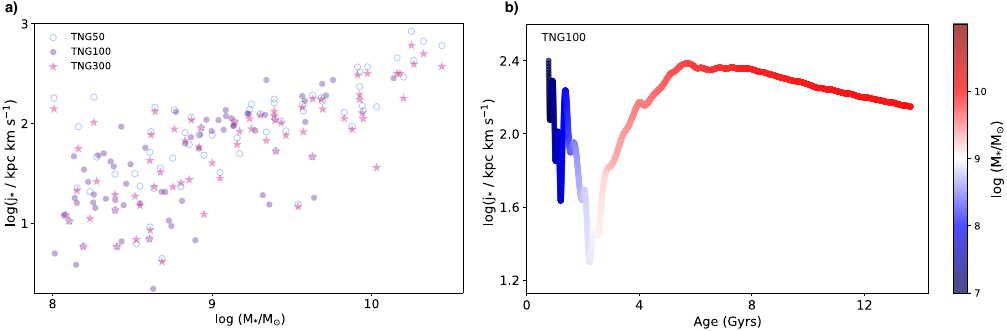}
   \caption{Consistency with TNG100 and TNG300 as a test of potential resolution effects. a) shows the $j$-M plot for a random sample of galaxies in TNG50 (blue), TNG100 (purple) and TNG300 (magenta). The distribution of galaxies within the $j$-M plane is consistent, including the increasing scatter towards lower $j$ values below $10^{9}\,\rm M_{*} (M_{\odot})$. b) shows an example of a galaxy within TNG100 experiencing a rapid increase in $j$ just below $10^{9}\,\rm M_{*} (M_{\odot})$, also consistent with TNG50 despite the lower resolution of TNG100. This suggests that the rapid jump within the $10^{8}$--$10^{9}$ M$_{\odot}$ mass range is more likely to be a physical effect rather than a resolution effect.}
 \label{resolution}
\end{figure*}

\label{lastpage}
\end{document}